\documentclass[aps,prd,twocolumn,superscriptaddress,runinaddress,floatfix,preprintnumbers,showpacs]{revtex4}

\usepackage{graphicx}%
\usepackage{dcolumn}% 
\usepackage{bm}% bold math
\usepackage{amsmath}
\usepackage{hyperref} 
\usepackage{xcolor}

\begin{document}

\preprint{ADP-09-09/T687}

\title{Isolating Excited States of the Nucleon in Lattice QCD}

%=========================================
\author{M.S. Mahbub}
\affiliation{Special Research Centre for the Subatomic Structure of Matter, Adelaide, South Australia 5005, Australia, \\ and Department of Physics, University of Adelaide, South Australia 5005, Australia.}\affiliation{Department of Physics, Rajshahi University, Rajshahi 6205, Bangladesh.}
\author{Alan $\acute{\rm{O}}$ Cais}
\author{Waseem Kamleh}
\author{B.G. Lasscock}
\author{Derek B. Leinweber}
\author{Anthony G. Williams}
\affiliation{Special Research Centre for the Subatomic Structure of Matter, Adelaide, South Australia 5005, Australia, \\ and Department of Physics, University of Adelaide, South Australia 5005, Australia.}

\collaboration{CSSM Collaboration}

\begin{abstract}

We discuss a robust projection method for the extraction of
excited-state masses of the nucleon from a matrix of correlation
functions.  To illustrate the algorithm in practice, we present
results for the positive parity excited states of the nucleon in
quenched QCD.  Using eigenvectors obtained via the variational method,
we construct an eigenstate-projected correlation function amenable to
standard analysis techniques.  The method displays its utility when
comparing results from the fit of the projected correlation function
with those obtained from the eigenvalues of the variational
method. Standard nucleon interpolators are considered, with $2\times
2$ and $3\times 3$ correlation matrix analyses presented using various
combinations of source-smeared, sink-smeared and smeared-smeared
correlation functions.
Using these new robust methods, we observe a systematic dependency of
the extracted nucleon excited-state masses on source- and sink-smearing levels.
To the best of our knowledge, this is the first clear indication that 
a correlation matrix of standard nucleon interpolators is insufficient
to isolate the eigenstates of QCD.

\end{abstract}

\pacs{11.15.Ha,12.38.Gc,12.38.-t}

\maketitle

\section{ \label{sec1:intro}Introduction}

Lattice QCD provides a non-perturbative tool to explore many
properties of hadrons from first principles. In the case of the hadron
mass spectrum there are well developed methods to compute the mass
spectra. However, while the extraction of the ground states of the
hadron spectrum is a well understood problem, and has provided
impressive agreement with experimental results ~\cite{Burch:2006cc},
the excited states still prove a significant challenge. The
Euclidean-time correlation function provides access to a tower of
states since it is a sum of decaying exponentials with the masses of
the states in the exponents. The ground state mass, being the lowest
energy state and thereby having the slowest decay rate, is obtained by
analysis of the large time behaviour of this function. The excited
states, however, belong to the sub-leading exponentials of the
two-point correlation function. Extracting excited states masses from
these sub-leading exponents is difficult as the correlation
functions decay quickly and the signal to noise ratio 
deteriorates more rapidly.

One of the long-standing puzzles in hadron spectroscopy has been the
low mass of the first positive parity, $J^{P}={\frac{1}{2}}^{+}$,
excitation of the nucleon, known as the Roper resonance $N^{*}$(1440
MeV). In constituent or valence quark models with harmonic oscillator
potentials, the lowest-lying odd parity states naturally occurs below
the $N={\frac{1}{2}}^{+}$ state (with principal quantum number $N=2$)
~\cite{Isgur:1977ef,Isgur:1978wd} whereas, in nature the Roper
resonance is almost 100 MeV below the $N={\frac{1}{2}}^{-}$(1535 MeV)
state. Similar difficulties in the level orderings appear for the
$J^{P}={\frac{3}{2}}^{+} \Delta^{\ast}(1600)$ and ${\frac{1}{2}}^{+}
\Sigma^{\ast} (1690)$ resonances, which have led to the speculation
that the Roper resonance may be more appropriately viewed as a hybrid
baryon state with explicitly excited glue field configurations
~\cite{Li:1991yba,Carlson:1991tg} or as a breathing mode of the
ground state ~\cite{Guichon:1985ny} or states which can be described
in terms of meson-baryon dynamics alone ~\cite{Krehl:1999km}. The
first detailed analysis of the positive parity excitation of nucleon
was performed in Ref.~\cite{Leinweber:1994nm} using Wilson fermions
and an operator product expansion spectral ansatz. Since then several
attempts have been made to address these issues in the lattice
framework
~\cite{Lee:1998cx,Gockeler:2001db,Sasaki:2001nf,Melnitchouk:2002eg,Edwards:2003cd,Lee:2002gn,Mathur:2003zf,Sasaki:2003xc},
but in many cases no potential identification of the Roper state has
been
made~\cite{Lee:1998cx,Gockeler:2001db,Sasaki:2001nf,Melnitchouk:2002eg,Edwards:2003cd}. Recently
however, in the analysis of ~\cite{Lee:2002gn,Mathur:2003zf,Sasaki:2005ap}, a
low-lying Roper state has been observed by using advanced fitting
techniques ~\cite{Lepage:2001ym,Morningstar:2001je} based on Bayesian
priors. Significant finite volume effects on the first positive parity
$\rm{N}^{{\frac{1}{2}}^{+}}$ state have been observed in
Refs.~\cite{Sasaki:2002sj,Sasaki:2003xc,Sasaki:2003vt} using the
Maximum Entropy Method
~\cite{Nakahara:1999vy,Allton:2002mr,Asakawa:2000tr,Lepage:2001ym,Morningstar:2001je}. Here,
we use another state-of-the-art approach, namely `the variational
method' ~\cite{Michael:1985ne,Luscher:1990ck,McNeile:2000xx}, which is
based on the correlation matrix analysis and has been used quite
extensively in
Refs.~\cite{Allton:1993wc,Sasaki:2001nf,McNeile:2000xx,Melnitchouk:2002eg,Zanotti:2003fx,Hedditch:2003zx,Lasscock:2007ce,Brommel:2003jm,Burch:2004he,Burch:2004zx,Burch:2005vn,Lasscock:2005tt,Burch:2005wd,Burch:2006dg,Burch:2006cc,Burch:2005md,Burch:2005qf,Basak:2006ki,Basak:2007kj,Mahbub:2009aa}
with the first analysis of the nucleon performed by
  Sasaki  \textit{et al} ~\cite{Sasaki:2001nf}. Though
the ground state mass of the nucleon has been described successfully,
an unambiguous determination of the Roper state has not been
successful to date with this method, though significant amounts of
research have been carried out in Ref.~\cite{Sasaki:2001nf,Allton:1993wc}, the CSSM
Lattice Collaboration
~\cite{Melnitchouk:2002eg,Zanotti:2003fx,Lasscock:2007ce}, the BGR
~\cite{Brommel:2003jm,Burch:2004he,Burch:2004zx,Burch:2005vn,Burch:2006cc}
collaboration and in Refs.~\cite{Basak:2006ki,Basak:2007kj}.

In this paper, we discuss an analysis method to extract masses of the
nucleon from the correlation functions using a variational
analysis. Employing standard interpolating operators $\chi_{1}$,
$\chi_{2}$ and $\chi_{4}$, we discuss the method for $2\times 2$ and
$3\times 3$ correlation matrices with the point and a range of sweeps
of Gaussian smearing ~\cite{Gusken:1989qx} at the source, sink and at
both source and sink. This analysis shows for the first time (and
despite the fact that people have been using source smeared
correlation functions for quite long time
~\cite{Melnitchouk:2002eg,Burch:2004he,Burch:2006dg,Burch:2006cc,Brommel:2003jm,Lasscock:2005tt,Lasscock:2005kx,Lasscock:2007ce,Alexandrou:2007qq,Alexandrou:2008tn,Edwards:2003cd,Richards:2001bx,Hedditch:2003dm,Hedditch:2003zx})
that, unexpectedly, the excited states of the nucleon are smearing
dependent. This analysis indicates that significant caution should be
taken when employing a particular level of smearing. To ensure the
maximal independence of our results on human input and minimal errors,
we construct a `robot' algorithm, governed by defined fitting
criteria, that automatically performs a standardised fitting
procedure. We present here results from this algorithm and also those
obtained from the eigenvalues to provide confidence in the extraction
of the nucleon mass spectrum.

This paper is arranged as follows:
Section~\ref{section:extraction_of_mass} contains the general
description of the extraction of masses with the introduction of
different nucleon interpolating fields. The lattice details are in
Section~\ref{section:lattice_details}, the analysis method is
presented in Section~\ref{section:analysis_and_discussion}, and
conclusions are presented in Section~\ref{section:conclusion}.     
%===============================================================
%
%
\section{Mass of Hadrons}
\label{section:extraction_of_mass}
The masses of hadrons are extracted from two-point correlation
functions using operators chosen to have overlap with desired
states. Let us consider a baryon state B of spin half, if we suppress
Dirac indices a two point function can be written as,  
%
%\vspace{-0.25cm}
\begin{align}
 {G_{ij}(t,\vec p)}&=\sum_{\vec x}e^{-i{\vec p}.{\vec x}}\langle{\Omega}\vert T \{ \chi_i (x)\bar\chi_j(0)\} \vert{\Omega}\rangle. 
\label{sec:mass:first_eqn}
\end{align}
The operator $\chi_j(0)$ creates states from the vacuum at space-time
point $0$ and, following the evolution of the states in time $t$, the
states are destroyed by the operator $\chi_{i}(x)$ at point $x$. $T$
stands for the time ordered product of operators. 
Having a complete set of momentum eigenstates requires that,
%\vspace{-0.25cm}
\begin{align}
\sum_{B,\vec {p'},s}\vert{B,\vec {p'},s}\rangle\langle{B,\vec {p'},s}\vert &=I,
\label{sec:mass:completeness_eqn}
\end{align}
where $B$ can include multi-particle states. The substitution of Eq.(\ref{sec:mass:completeness_eqn}) into the Eq.(\ref{sec:mass:first_eqn}) yields,
%
%\vspace{-0.25cm}
\begin{align}
{G_{ij}(t,\vec p)} &=\sum_{\vec x}\sum_{B,\vec {p'},s}e^{-i{\vec p}.{\vec x}}\langle {\Omega}\vert\chi_i(x)\vert{B,\vec {p'},s}\rangle\langle{B,\vec {p'},s}\vert\bar\chi_j(0)\vert {\Omega}\rangle.\label{sec:mass:MinkowskiCorrelationFunction_eqn1}
\end{align}
We can express the operator $\chi_{i}(x)$ as
%\vspace{-0.25cm}
\begin{align}
\chi_{i}(x) &= e^{iP.x}\chi_{i}(0)e^{-iP.x},
\end{align}
where, $P^{\mu}=P=(H,\vec{P})$ and $\vec{P}$ is the momentum operator whose eigenvalue is the total momentum of the system. Eq.(\ref{sec:mass:MinkowskiCorrelationFunction_eqn1}) can now be written as, 
%
%\vspace{-0.25cm}
\begin{widetext}
\begin{align}
{G_{ij}(t,\vec p)} &=\sum_{\vec x}\sum_{B,\vec{ p'},s}e^{-i{\vec p}.{\vec x}}\langle {\Omega}\vert{e^{iPx}\chi_{i}(0)e^{-iPx}}\vert{B,\vec {p'},s}\rangle\langle{B,\vec {p'},s}\vert\bar\chi_{j}(0)\vert {\Omega}\rangle \nonumber \\
&=\sum_{\vec x}\sum_{B,\vec {p'},s}e^{-iE_{B}t}e^{-i{\vec x}.({\vec p}-\vec{ p'})}\langle {\Omega}\vert\chi_{i}(0)\vert{B,\vec {p'},s}\rangle\langle{B,\vec {p'},s}\vert\bar\chi_{j}(0)\vert {\Omega}\rangle. 
\end{align}
\end{widetext}
As we move from Minkowski space to Euclidean space, the time $t\rightarrow {-it}$ and the above equation then can be written as,
%
%\vspace{-0.25cm}
\begin{align}
{G_{ij}(t,\vec p)} &=\sum_{B,\vec {p'},s}e^{-E_{B}t}\delta_{\vec p,\vec {p'}}\langle {\Omega}\vert\chi_{i}(0)\vert{B,\vec {p'},s}\rangle\langle{B,\vec {p'},s}\vert\bar\chi_{j}(0)\vert {\Omega}\rangle\nonumber \\
&=\sum_{B}\sum_{s}e^{-E_{B}t}\langle {\Omega}\vert\chi_{i}(0)\vert{B,{\vec p},s}\rangle\langle{B,{\vec p},s}\vert\bar\chi_{j}(0)\vert {\Omega}\rangle.
\label{sec:mass:EucleadianCorrelationFunction_eqn1}
\end{align}
The overlap of the interpolating fields $\chi(0)$ and ${\bar\chi}(0)$ with positive and negative parity baryon states $\vert {B^{\pm}}\rangle$ can be parametrized by a complex quantity called the coupling strength, $\lambda_{B^{\pm}}$, which can be defined for positive parity states by
%
%\vspace{-0.25cm}
\begin{align}
\langle{\Omega}\vert\chi(0)\vert {B^{+}},\vec p,s\rangle &=\lambda_{B^{+}}\sqrt {\frac{M_{B^{+}}}{E_{B^{+}}}}u_{B^{+}}({\vec p},s),
\end{align}
%
%\vspace{-0.5cm}
\begin{align}
\langle B^{+},\vec p,s\vert\bar{\chi}(0)\vert {\Omega}\rangle &=\bar\lambda_{B^{+}}\sqrt {\frac{M_{B^{+}}}{E_{B^{+}}}}{{\bar u}_{B^{+}}}({\vec p},s).
\end{align}
For the negative parity states the definition is
%
%\vspace{-0.25cm}
\begin{align}
\langle{\Omega}\vert\chi(0)\vert {B^{-}},\vec p,s\rangle &=\lambda_{B^{-}}\sqrt {\frac{M_{B^{-}}}{E_{B^{-}}}}\gamma_5{u_{B^{-}}({\vec p},s)},
\end{align}
%
%\vspace{-0.5cm}
\begin{align}
\langle B^{-},\vec p,s\vert\bar{\chi}(0)\vert {\Omega}\rangle &=-\bar\lambda_{B^{-}}\sqrt {\frac{M_{B^{-}}}{E_{B^{-}}}}{{\bar u}_{B^{-}}}({\vec p},s)\gamma_5.
\end{align}
Here, $\lambda_{B^{\pm}} $ and ${\bar\lambda}_{B^{\pm}}$ are the couplings of the interpolating functions at the sink and the source respectively and $M_{B^{\pm}}$ is the mass of the state $B^{\pm}$. ${E_{B^{\pm}}}$ is the energy of the state $B^{\pm}$, where ${E_{B^{\pm}}} = \sqrt{M^{2}_{B^{\pm}}+{\vec p}^2}$, and $u_{B^{\pm}}(\vec p,s)$ and ${{\bar u}_{B^{\pm}}}(\vec p,s)$ are the Dirac spinors, 
%
%\vspace{-0.25cm}
\begin{align}
{\bar u}^{\alpha}_{B^{\pm}}(\vec p,s) {u^{\beta}_{B^{\pm}}}(\vec p,s) &= \delta{^{\alpha \beta}}.
\end{align}
Thus, Eq.(\ref{sec:mass:EucleadianCorrelationFunction_eqn1}) contains a projection operator $\Gamma_{\pm}=\sum_{s} {u^{\beta}_{B^{\pm}}}(\vec p,s){\bar u}^{\alpha}_{B^{\pm}}(\vec p,s)$, through which the contributions to the even and odd parity states from the correlation function can be obtained. For positive parity, this can be expressed as, 
%
%\vspace{-0.25cm}
\begin{align} 
\sum_{s} {u^{\beta}_{B^{+}}}(\vec p,s){\bar u}^{\alpha}_{B^{+}}(\vec p,s) &=\frac{\gamma .p + M_{B^{+}}}{2{E_{B^{+}}}},
\end{align}

and for the negative parity,
%\vspace{-0.25cm}
\begin{align} 
\gamma_{5}\left(\sum_{s} {u^{\beta}_{B^{-}}}(\vec p,s){\bar u}^{\alpha}_{B^{-}}(\vec p,s)\right)\gamma_{5} &=\frac{-\gamma .p + M_{B^{-}}}{2{E_{B^{-}}}}.
\end{align}
By substituting the above Eqs. for the positive and negative parity states in Eq.(\ref{sec:mass:EucleadianCorrelationFunction_eqn1}) we obtain,
%
%\vspace{-0.25cm}
%\begin{widetext}
\begin{align}
{\cal{G}}_{ij}(t,\vec p) &=\sum_{B^{+}}\lambda_{B^{+}}\bar\lambda_{B^{+}}e^{{-E_{B^{+}}}t} {\frac{\gamma .p_{B^{+}} + M_{B^{+}}}{2E_{B^{+}}}} \nonumber \\
              & +\sum_{B^{-}}\lambda_{B^{-}}\bar\lambda_{B^{-}}e^{{-E_{B^{-}}}t} {\frac{-\gamma .p_{B^{-}} + M_{B^{-}}}{2E_{B^{-}}}}.\label{sec:mass:FinalCorrelationFunction_eqn}
\end{align}
%
%\end{widetext}
At momentum $\vec p=\vec 0$, $E_{B^{\pm}}=M_{B^{\pm}}$, and a parity projection operator $\Gamma_{\pm}$ can be introduced,
%
%\vspace{-0.25cm}
\begin{align}
\Gamma_{\pm} &= \frac{1}{2}(1\pm \gamma_0).
\end{align}
 We can isolate the masses of the even and odd parity states by taking the trace of $\cal{G}$ with the operators $\Gamma_{+}$ and $\Gamma_{-}$. The positive parity state propagates through the (1,1) and (2,2) elements of the Dirac matrix, whereas, negative parity state propagates through the (3,3) and (4,4) elements.

The correlation function for positive and negative parity states can then be written as,
%
%\vspace{-0.25cm}
\begin{align}
G_{ij}^{\pm}(t,\vec 0) &= {\rm{Tr}}_{\rm sp}[\Gamma_{\pm}{\cal{G}}_{ij}(t,\vec 0)]\nonumber \\
&= \sum_{B^{\pm}}\lambda_{i}^{\pm}\bar\lambda_{j}^{\pm}e^{{-M_{B^{\pm}}}t}.
\end{align}
The correlation function contains a superposition of states. The mass of the lowest state, $M_{0^{\pm}}$ can be extracted at large $t$ where the contributions from all other states are suppressed,
%
%\vspace{-0.25cm}
\begin{align}
G_{ij}^{\pm}(t,\vec 0) & \stackrel{t \rightarrow \infty}{=} \lambda_{i0}^{\pm}\bar\lambda_{j0}^{\pm}e^{{-M_{0^{\pm}}}t}.
\end{align}

The source smearing~\cite{Gusken:1989qx} technique is applied to
increase the overlap of the interpolators with the lower lying
states. A fixed boundary condition in the time direction is applied
for the fermions by setting $U_t(\vec x,N_t)=0\,\forall\,{\vec x}$ in
the hopping terms of the fermion action with periodic boundary
conditions imposed in the spatial directions. Gauge invariant Gaussian
smearing ~\cite{Gusken:1989qx} in the spatial dimensions is applied
through an iterative process. The smearing procedure is: 
%
%\vspace{-0.25cm}
\begin{align}
\psi_{i}(x,t) &=\sum_{x'}F(x,x')\psi_{i-1}(x',t),
\end{align}
%\noindent
where,
%
%\vspace{-0.25cm}

\begin{align}
F(x,x') &= {(1-\alpha)}\delta_{x,x'}+\frac{\alpha}{6}\sum_{\mu=1}^{3}[U_{\mu}(x)\delta_{x',x+\hat\mu} \nonumber \\
        & +U_{\mu}^{\dagger}(x-\hat\mu)\delta_{x',x-\hat\mu}],
\end{align}

%\noindent
where the parameter $\alpha=0.7$ is used in our calculation. After repeating the procedures $N_{\rm sm}$ times on a point source the resulting smeared fermion field is,
\begin{align}
\psi_{N_{\rm sm}}(x,t) &=\sum_{x'}F^{N_{sm}}(x,x')\psi_{0}(x',t).
\end{align}

%=============================================================================

%\noindent
The extraction of the ground state mass is done
straightforwardly. However access to the excited state masses requires
additional effort. Here we consider the variational method
~\cite{Michael:1985ne,Luscher:1990ck,McNeile:2000xx}. The variational
method requires the cross correlation of operators so that the
operator space can be diagonalised and the excited state masses
extracted from the exponential nature of the diagonalised basis. To
access $N$ states of the spectrum, one requires a minimum of $N$
interpolators. Traditionally, only a few interpolators are considered
providing access to a small number of states of the desired channel. 

The parity projected two point correlation function matrix for
$\vec{p} =0$ can be written as, 
%
%\vspace{-0.25cm}
\begin{align}
G_{ij}(t) &= (\sum_{\vec x}{\rm Tr}_{\rm sp}\{ \Gamma_{\pm}\langle\Omega\vert\chi_{i}(x)\bar\chi_{j}(0)\vert\Omega\rangle\}) \\
          &=\sum_{\alpha=0}^{N-1}\lambda_{i}^{\alpha}\bar\lambda_{j}^{\alpha}e^{-m_{\alpha}t}.
\end{align}
Here, $\lambda_{i}^{\alpha}$ and $\bar\lambda_{j}^{\alpha}$ are the
couplings of interpolators $\chi_{i}$ and  $\bar\chi_{j}$ at the sink
and source respectively to eigenstates $\alpha=0, \cdots
,(N-1)$. $m_{\alpha}$ is the mass of the state $\alpha$. The $N$
interpolators have the same quantum numbers and provide an
$N$-dimensional basis upon which to describe the states. Using this
basis we aim to construct $N$ independent interpolating source and
sink fields which isolate $N$ baryon states $\vert B_{\alpha}\rangle,$
{\it i.e.} 
%\vspace{-0.25cm}
\begin{align}
{\bar\phi}^{\alpha} &=\sum_{i=1}^{N}u_{i}^{\alpha}{\bar\chi}_{i},
\end{align}   
\vspace{-0.60cm}
\begin{align}
{\phi}^{\alpha} &=\sum_{i=1}^{N}v_{i}^{\alpha}{\chi}_{i},
\end{align} 
%
%\noindent
such that,
%\vspace{-0.25cm}
\begin{align}
\langle{B_{\beta},p,s}\vert {\bar\phi}^{\alpha}\vert\Omega\rangle &= \delta_{\alpha\beta}{\bar{z}}^{\alpha}\bar{u}(\alpha,p,s),
\end{align}
\vspace{-0.60cm}
\begin{align}
\langle\Omega\vert{\phi}^{\alpha}\vert B_{\beta},p,s\rangle &= \delta_{\alpha\beta}{z}^{\alpha}u(\alpha,p,s),
\end{align}
%
%\noindent
where $z^{\alpha}$ and ${\bar{z}}^{\alpha}$ are the coupling strengths
of $\phi^{\alpha}$ and ${\bar\phi}^{\alpha}$ to the state $\vert
B_{\alpha}\rangle$. Consider a real eigenvector $u_{j}^{\alpha}$ which
operates on the correlation matrix $G_{ij}(t)$ from right, one can
obtain ~\cite{Melnitchouk:2002eg}, 
%\vspace{-0.25cm}
%
\begin{align}
G_{ij}(t)u_{j}^{\alpha} &=(\sum_{\vec x}{\rm Tr}_{\rm sp}\{ \Gamma_{\pm}\langle\Omega\vert\chi_{i}\bar\chi_{j}\vert\Omega\rangle\})u_{j}^{\alpha} \nonumber \\
& = \lambda_{i}^{\alpha}\bar{z}^{\alpha}e^{-m_{\alpha}t}.
\end{align}
For notational convenience, in the remainder of the discussion the
repeated indices $i,j,k$ are to be understood as being summed over,
whereas, $\alpha$, which stands for a particular state, is not. Since
the only $t$ dependence comes from the exponential term, we can write
a recurrence relation at time $(t+\triangle t)$ as, 
%\vspace{-0.25cm}
\begin{align}
G_{ij}(t+\triangle t)u_{j}^{\alpha} & = e^{-m_{\alpha}\triangle t} G_{ij}(t)u_{j}^{\alpha}.
\end{align}  
If we multiply the above equation by $[G_{ij}(t)]^{-1}$ from the left we get,
%
%\vspace{-0.25cm}
\begin{align}
[(G(t))^{-1}G(t+\triangle t)]u^{\alpha} & = e^{-m_{\alpha}\triangle t}u^{\alpha} \nonumber \\
& = c^{\alpha}u^{\alpha}.
\label{eqn:right_eigenvalue_equation}
\end{align} 
This is an eigenvalue equation for eigenvector $u^{\alpha}$ with eigenvalue $c^{\alpha}=e^{-m_{\alpha}\triangle t}$. We can also solve the left eigenvalue equation to recover the $v^{\alpha}$ eigenvector,
%\vspace{-0.25cm}
\begin{align}
v_{i}^{\alpha}G_{ij}(t+\triangle t) & = e^{-m_{\alpha}\triangle t}v_{i}^{\alpha}G_{ij}(t).
\end{align} 
Similarly,
%
%\vspace{-0.25cm}
\begin{align}
v^{\alpha}[G(t+\triangle t)(G(t))^{-1}] & = e^{-m_{\alpha}\triangle t}v^{\alpha}.
\label{eqn:left_eigenvalue_equation}
\end{align} 
The vectors $u_{j}^{\alpha}$ and $v_{i}^{\alpha}$  diagonalize the correlation matrix at time $t$ and $t+\triangle t$ making the projected correlation matrix,
%
%\vspace{-0.10cm}
\begin{align}
v_{i}^{\alpha}G_{ij}(t)u_{j}^{\beta} = \delta^{\alpha\beta}z^{\alpha}{\bar{z}}^{\beta}e^{-m_{\alpha}t}.
 \label{eqn:projected_cf} 
\end{align} 
The parity projected, eigenstate projected correlator, $v_{i}^{\alpha}G^{\pm}_{ij}(t)u_{j}^{\alpha} \equiv G_{\pm}^{\alpha}$ is then used to obtain masses of different states.
We construct the effective mass 
\begin{align}
M_{\rm eff}^{\alpha}(t) &= {\rm ln}\left(\frac{{G_{\pm}^{\alpha}}(t,\vec 0)}{G_{\pm}^{\alpha}(t+1,\vec 0)}\right)\nonumber \\
 & = M_{\pm}^{\alpha}.
 \label{eqn:efective_mass}
\end{align}
and apply standard analysis techniques as described in the following.
%=============================================================================
\section{Lattice details}
\label{section:lattice_details}
We use an ensemble of 200 quenched configurations with a lattice
volume of $16^{3}\times 32.$ Gauge field configurations are generated
by using the DBW2 gauge action
~\cite{Takaishi:1996xj,deForcrand:1999bi} and an
${\cal{O}}(a)$-improved FLIC fermion action ~\cite{Zanotti:2001yb} is
used to generate quark propagators. This action has excellent scaling
properties and provides near continuum results at finite lattice
spacing ~\cite{Zanotti:2004dr}. The lattice spacing is $a=0.1273$ fm,
as determined by the static quark potential, with the scale set with
the Sommer scale, $r_{o}=0.49$ fm ~\cite{Sommer:1993ce}. In the
irrelevant operators of the fermion action we apply four sweeps of
stout-link smearing to the gauge links to reduce the coupling with the
high frequency modes of the theory ~\cite{Morningstar:2003gk}. We use
the same method as in Ref.~\cite{Lasscock:2005kx} to determine fixed
boundary effects, and the effects are significant only after time
slice 25 in the present analysis. Beside point operators, various
sweeps (1, 3, 7, 12, 16, 26, 35, 48, 65 sweeps corresponding to rms
  radii, in lattice units, 0.6897, 1.0459, 1.5831, 2.0639, 2.3792,
  3.0284, 3.5237, 4.1868, 5.0067) of gauge invariant Gaussian smearing
~\cite{Gusken:1989qx} are applied at the source (at $t=4$)
and at the sink. This is to ensure a variety of overlaps of the
interpolators with the lower-lying states. The analysis is performed
on four different quark masses providing pion masses
${m_{\pi}}=\{0.797,0.729,0.641,0.541\}$ GeV. The error analysis is
performed using a second-order single elimination jackknife method,
where the ${\chi^{2}}/{\rm{dof}}$ is obtained via a covariance matrix
analysis method. We discuss our fitting method in the next
section.

The nucleon interpolators we consider in this analysis are,
\begin{align}
\chi_1(x) &= \epsilon^{abc}(u^{Ta}(x)C{\gamma_5}d^b(x))u^{c}(x) \\
\chi_2(x) &= \epsilon^{abc}(u^{Ta}(x)Cd^b(x)){\gamma_5}u^{c}(x) \\
\chi_4(x) &= \epsilon^{abc}(u^{Ta}(x)C{\gamma_5}{\gamma_4}d^b(x))u^{c}(x)
\end{align}
The $\chi_{1}$ and $\chi_{2}$ interpolators are used in
Refs.~\cite{Leinweber:1990dv,Sasaki:2001nf,Leinweber:1994nm}. The $\chi_{4}$
interpolator is considered as the time component of the $\chi_{3}$
interpolator used in
Refs.\cite{Brommel:2003jm,Zanotti:2003fx,Lasscock:2007ce}. We use the
Dirac representation of the gamma matrices in our analysis.

\section{Analysis and discussion}
\label{section:analysis_and_discussion}
\subsection{Correlation Matrix Analysis}
\begin{figure*}[tph] %[!hbp]
  \begin{center}
   $\begin{array}{c@{\hspace{0.15cm}}c}  
 \includegraphics [height=0.48\textwidth,width=0.34\textwidth,angle=90]{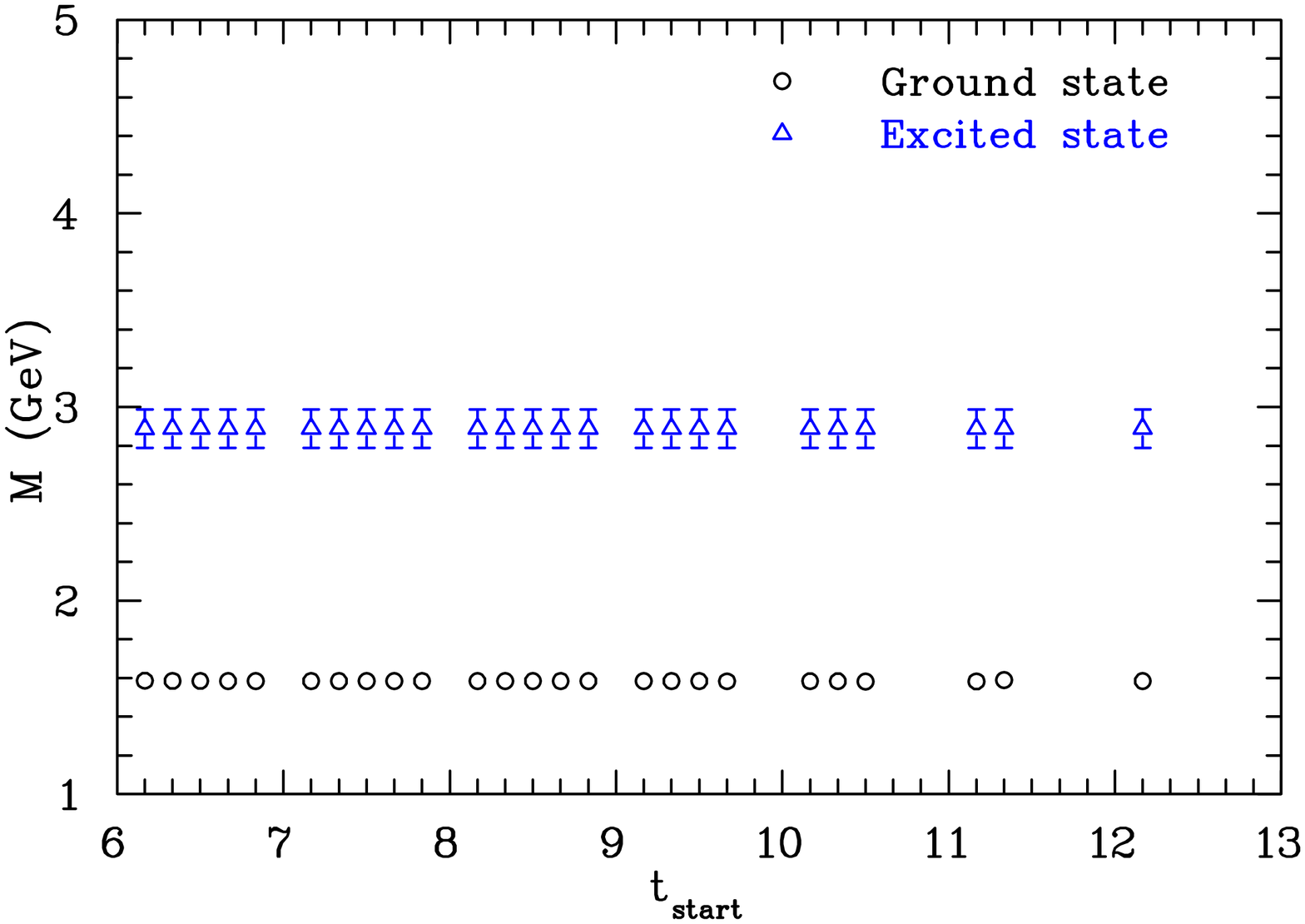} &
 \includegraphics [height=0.48\textwidth,width=0.34\textwidth,angle=90]{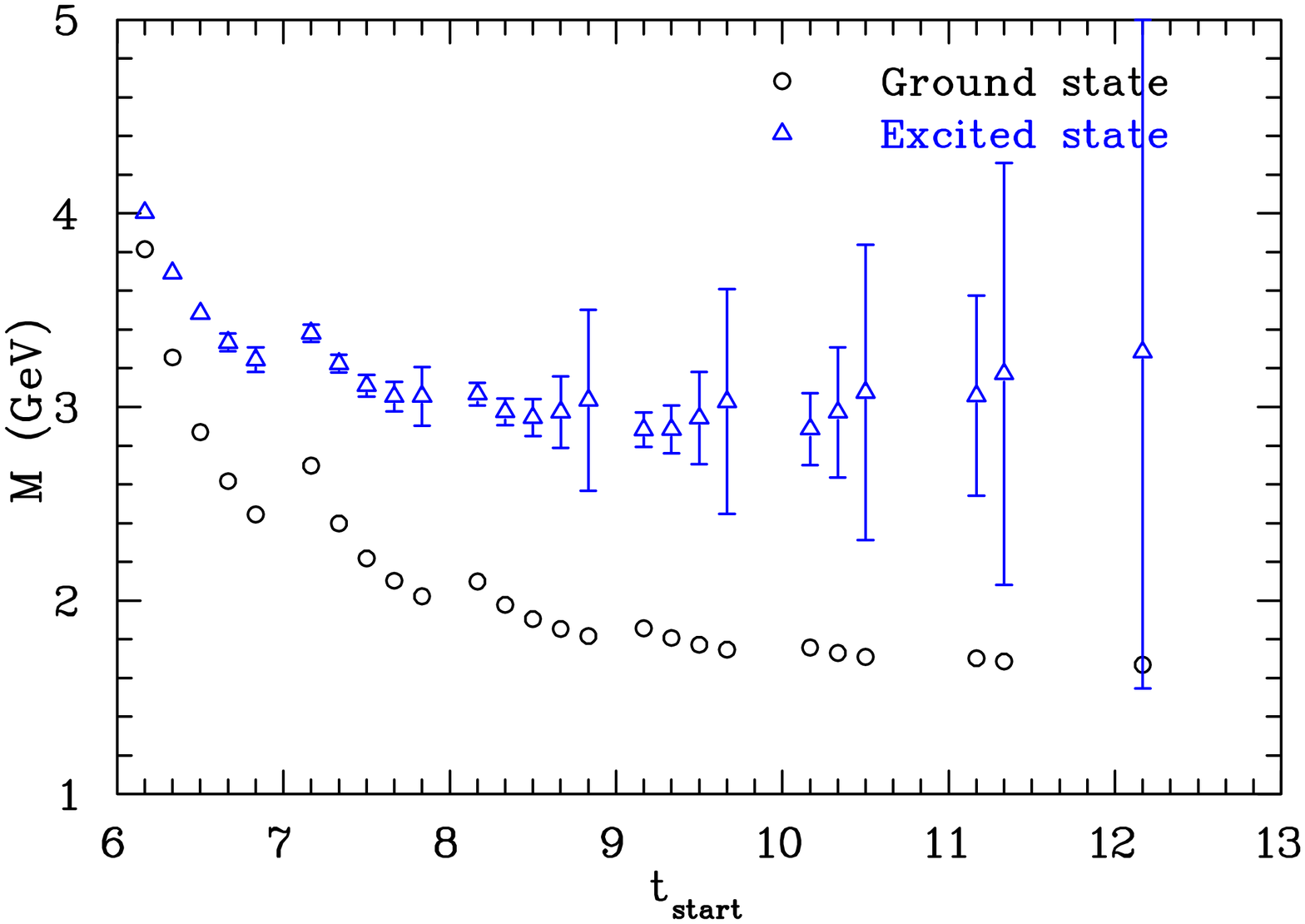}    
    \end{array}$
    \caption{(Color online). Mass of the nucleon
      ($N^{{\frac{1}{2}}^{+}}$) from the projected correlation
      function as shown in Eq.\ref{eqn:projected_cf} (left) and  from
      the eigenvalue (right) for a $2\times 2$ correlation matrix of
      $\chi_{1}$, $\chi_{2}$ interpolators. The figure corresponds to
      a pion mass of 797 MeV (heaviest) and for the \textbf{point
        source} to \textbf{point sink} correlation functions. Each pair
      of ground and excited states masses correspond to the
      diagonalization of the correlation matrix for each set of
      variational parameters $t_{\rm{start}}$ (shown in major tick marks) and
      $\triangle t$ (shown in minor tick marks). Here, the time $t$ as shown in Eqs.(\ref{eqn:right_eigenvalue_equation}, and \ref{eqn:left_eigenvalue_equation}) is called $t_{\rm{start}}$.}  
   \label{fig:mass_and_eig_for_ptpp_for_x1x2_h1Q}
  \end{center}
\end{figure*}
\begin{figure*}[tph] %[!hbp]
  \begin{center}
   $\begin{array}{c@{\hspace{0.15cm}}c}  
 \includegraphics [height=0.48\textwidth,width=0.34\textwidth,angle=90]{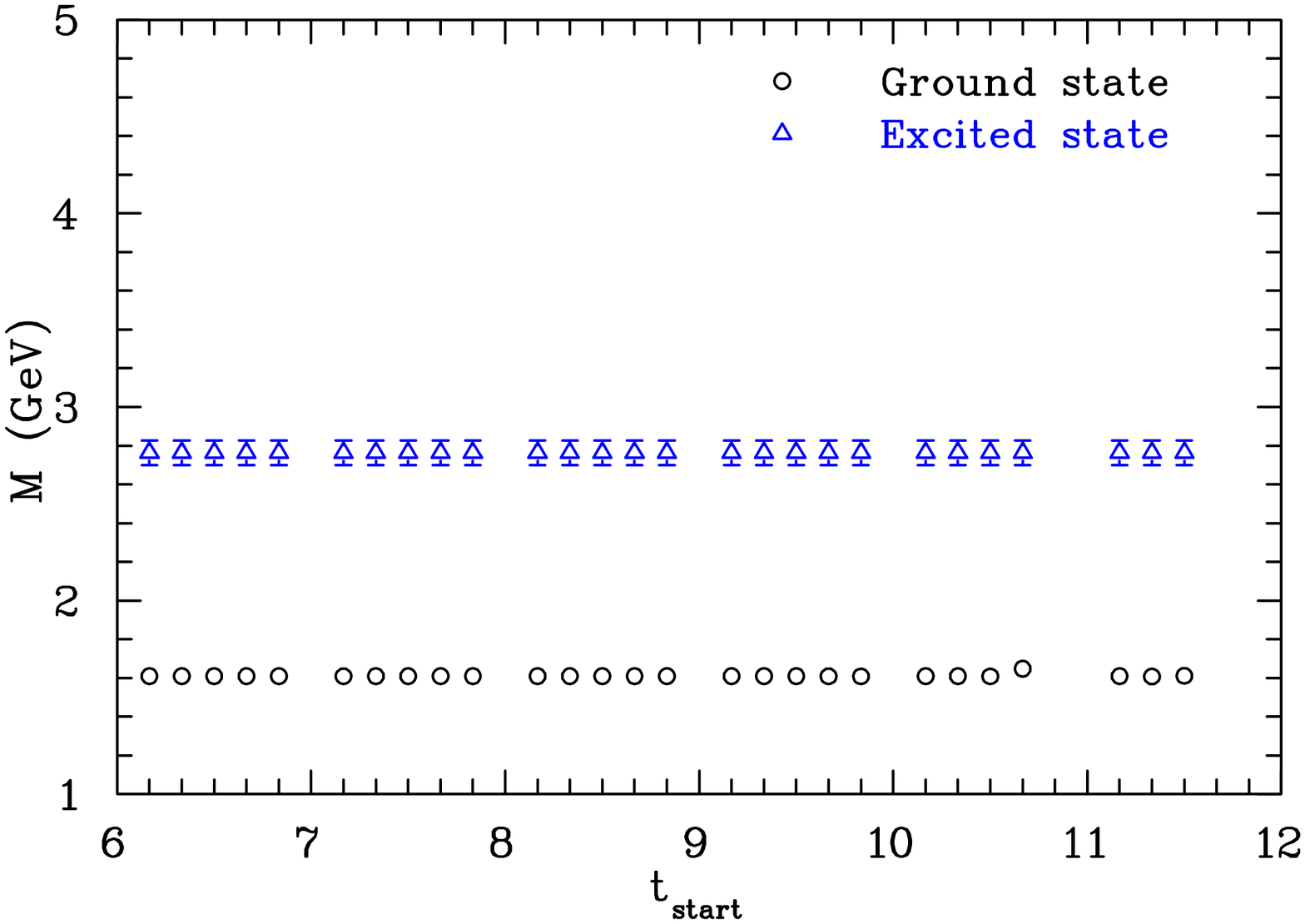} &
 \includegraphics [height=0.48\textwidth,width=0.34\textwidth,angle=90]{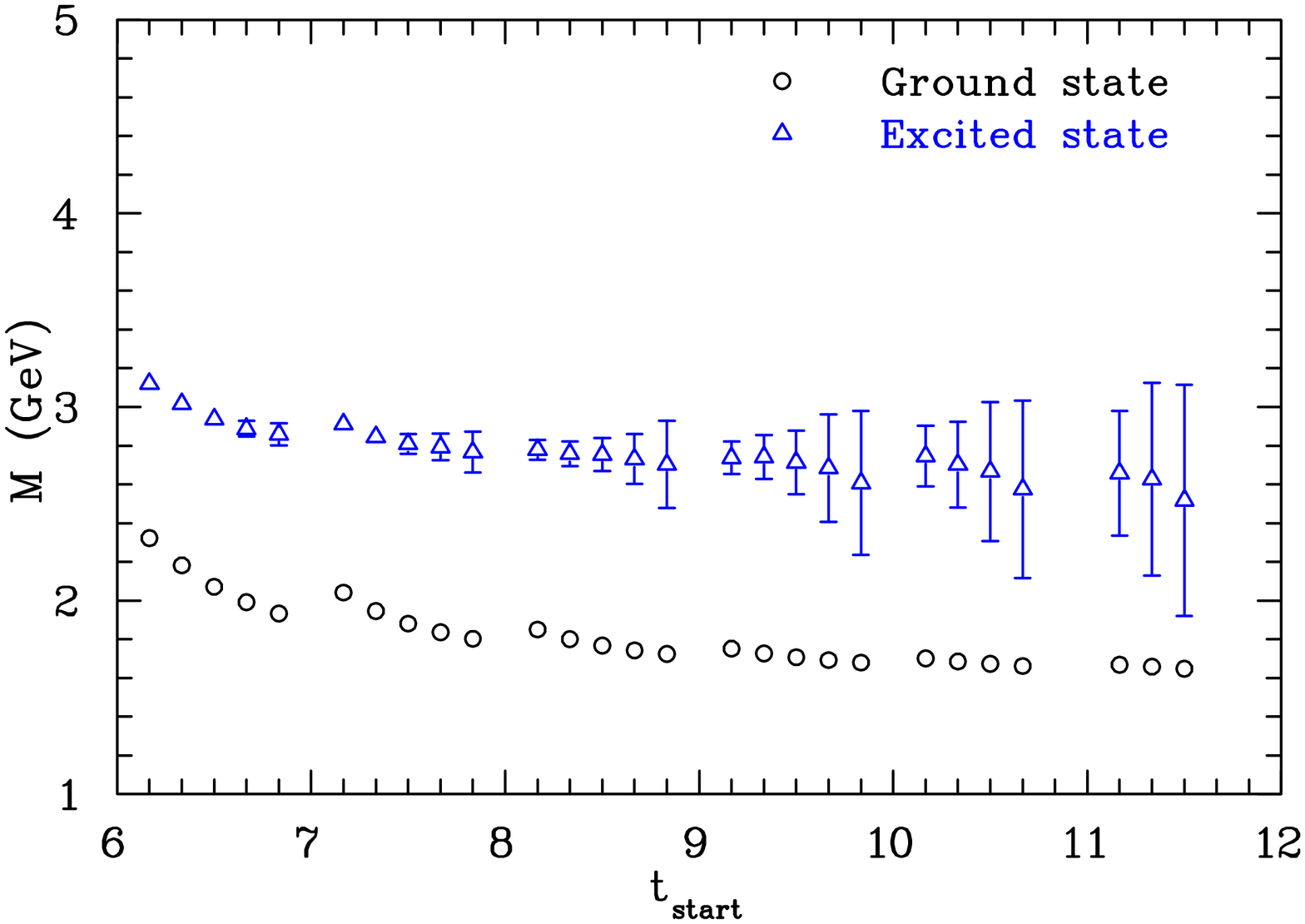}    
    \end{array}$
    \caption{(Color online). As in
      Fig.~\ref{fig:mass_and_eig_for_ptpp_for_x1x2_h1Q}, but for
      \textbf{source smeared} to \textbf{point sink} correlation
      functions with the number of Gaussian smearing sweeps set to 7, which
      corresponds to a smearing radius of 1.5831 in lattice units.}  
   \label{fig:mass_and_eig_for_so7_for_x1x2_h1Q}
  \end{center}
\end{figure*}
We begin this section by outlining the fitting method used in this
paper. The method is based on the maximisation of the Euclidean-time
fit window and the minimisation of the value of the
${\chi^{2}}/{\rm{dof}}$ commencing at the earliest time possible
~\cite{Cais:2008cx}.  At larger times the lighter states dominate the
correlation function. However, the signal to noise ratio decreases
rapidly, forcing increased errors in the results. Additionally,
ignoring smaller time slices may lead to eliminating important
information about excited states included in the two-point correlation
function ~\cite{Alexandrou:2008bp}.  Data at larger time slices for
excited states extracted using the variational method can be
contaminated by residual contributions from the lower lying states,
resulting in lower values of excited states mass. In consideration
of these points, in this analysis we use a preference towards earlier
times which have a high signal to noise ratio and are therefore
heavily constraining in the fit procedure. This allows us to isolate
the energy level from higher state contamination through the
${\chi^{2}}/{\rm{dof}}$ and simultaneously control the errors
potentially introduced at higher times (where the signal to noise
ratio is lower) by contamination from lower-lying states in the
varitiational analysis procedure. Hence, the inclusion of these early
times minimizes the error in the results while an
acceptable value of ${\chi^{2}}/{\rm{dof}}$ is maintained.\\
 We perform the analysis for $2\times 2$ and $3\times 3$ correlation
matrices via the variational method. For the quark masses considered herein, the interpolator $\chi_{1}$ has
better overlap with the lower-energy states ~\cite{Melnitchouk:2002eg}
and strongly couples to the nucleon ground state
~\cite{Sasaki:2001nf,Brommel:2003jm}, whereas the interpolator $\chi_{2}$ does not
have good overlap with the nucleon ground state and couples to the
higher energy state(s)
~\cite{Leinweber:1994nm,Sasaki:2001nf,Bowler:1984dh,Melnitchouk:2002eg}.  We have
found that the other interpolator, $\chi_{4}$, is very similar to
$\chi_{1}$ and also couples strongly to the ground state
~\cite{Brommel:2003jm}, which suggests that the $\chi_{1}$ and
$\chi_{4}$ operators are somewhat linearly dependent to each
other.  We call the start time, $t$, of the variational analysis
$t_{\rm{start}}$.  The diagonalization is accomplished for different
values of $t_{\rm{start}}$ with a few values of $\triangle t$, here
$\triangle t = 1-5$ for each $t_{\rm{start}}$.
\begin{figure*}[tph] %[!hbp]
  \begin{center}
   $\begin{array}{c@{\hspace{0.15cm}}c}  
 \includegraphics [height=0.48\textwidth,width=0.34\textwidth,angle=90]{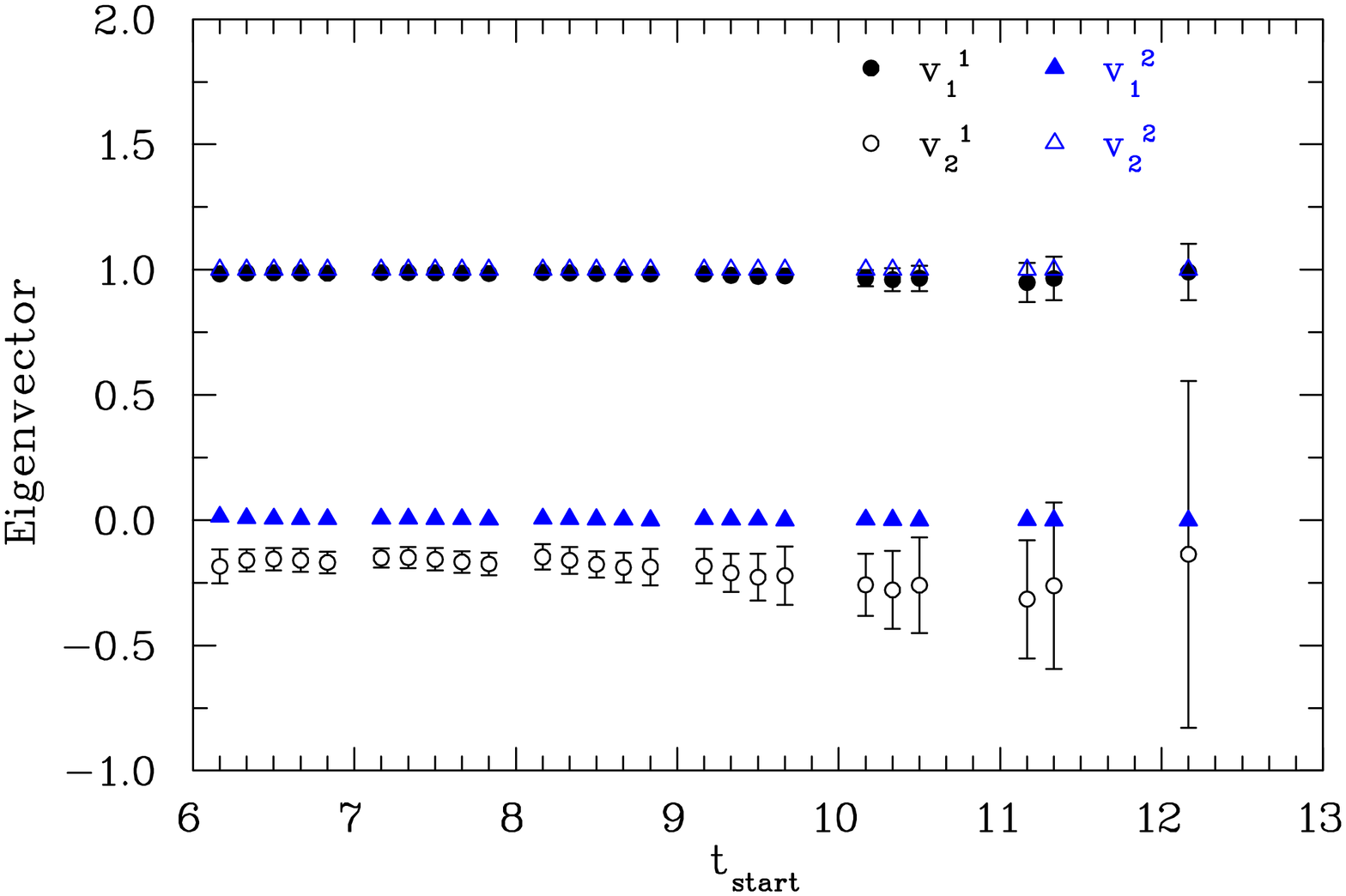} &
 \includegraphics [height=0.48\textwidth,width=0.34\textwidth,angle=90]{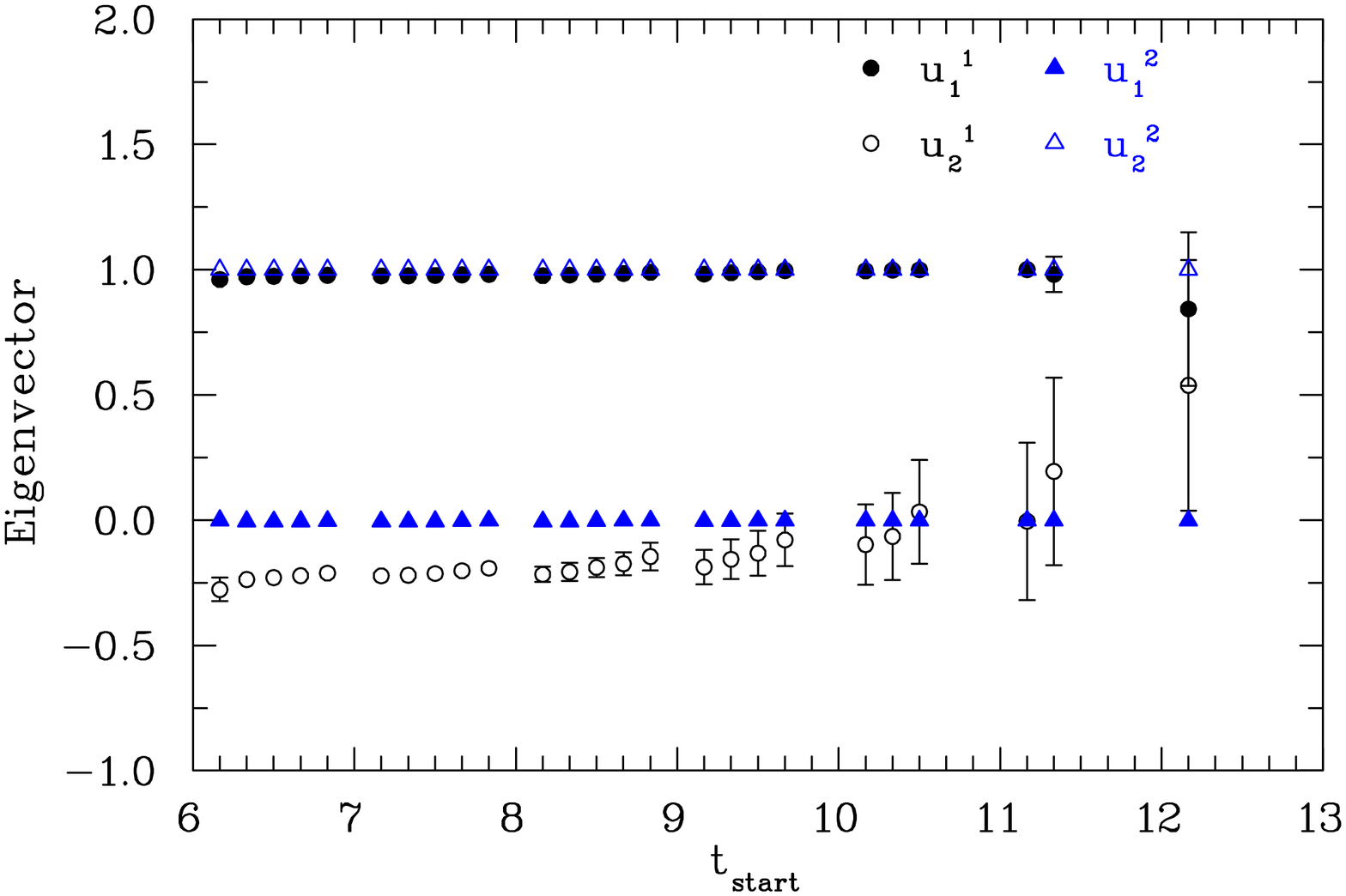}    
    \end{array}$
    \caption{(Color online). Eigenvector values for $v_{i}^{\alpha}$ (left) and
      $u_{i}^{\alpha}$ (right), as shown in
      Eq.(\ref{eqn:projected_cf}), for the correlation matrix analysis of
      Fig.~\ref{fig:mass_and_eig_for_ptpp_for_x1x2_h1Q}. The superscript $\alpha$ stands for the eigenstates while the subscript $i$ represents the interpolators. Here, for the 2x2 correlation matrix, $\alpha=1,2$ and $i=1,2$.}  
   \label{fig:evectors_for_ptpp_for_x1x2_h1Q}
  \end{center}
\end{figure*}

\begin{figure*}[tph] %[!hbp]
  \begin{center}
   $\begin{array}{c@{\hspace{0.15cm}}c}  
 \includegraphics [height=0.48\textwidth,width=0.34\textwidth,angle=90]{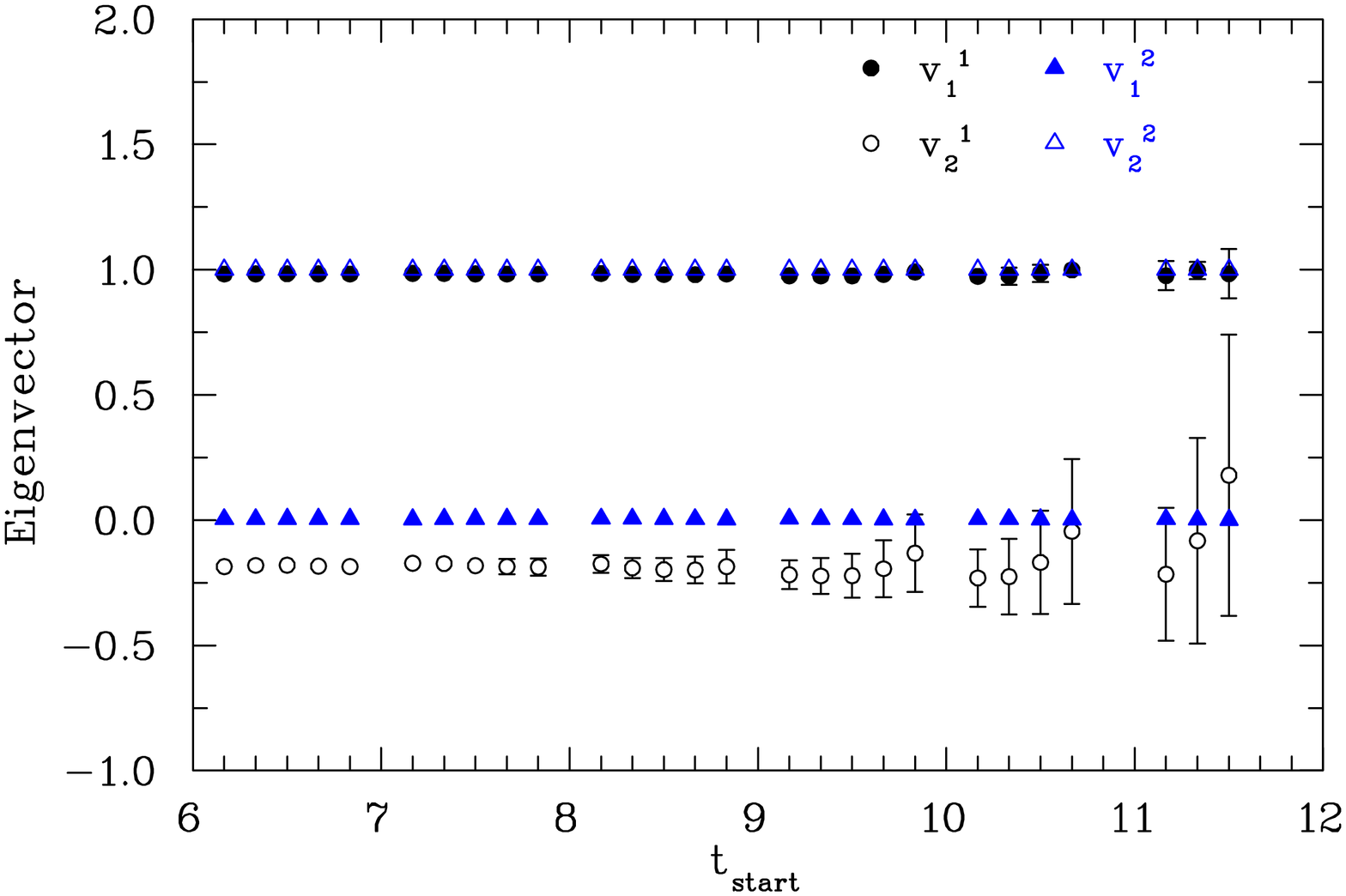} &
 \includegraphics [height=0.48\textwidth,width=0.34\textwidth,angle=90]{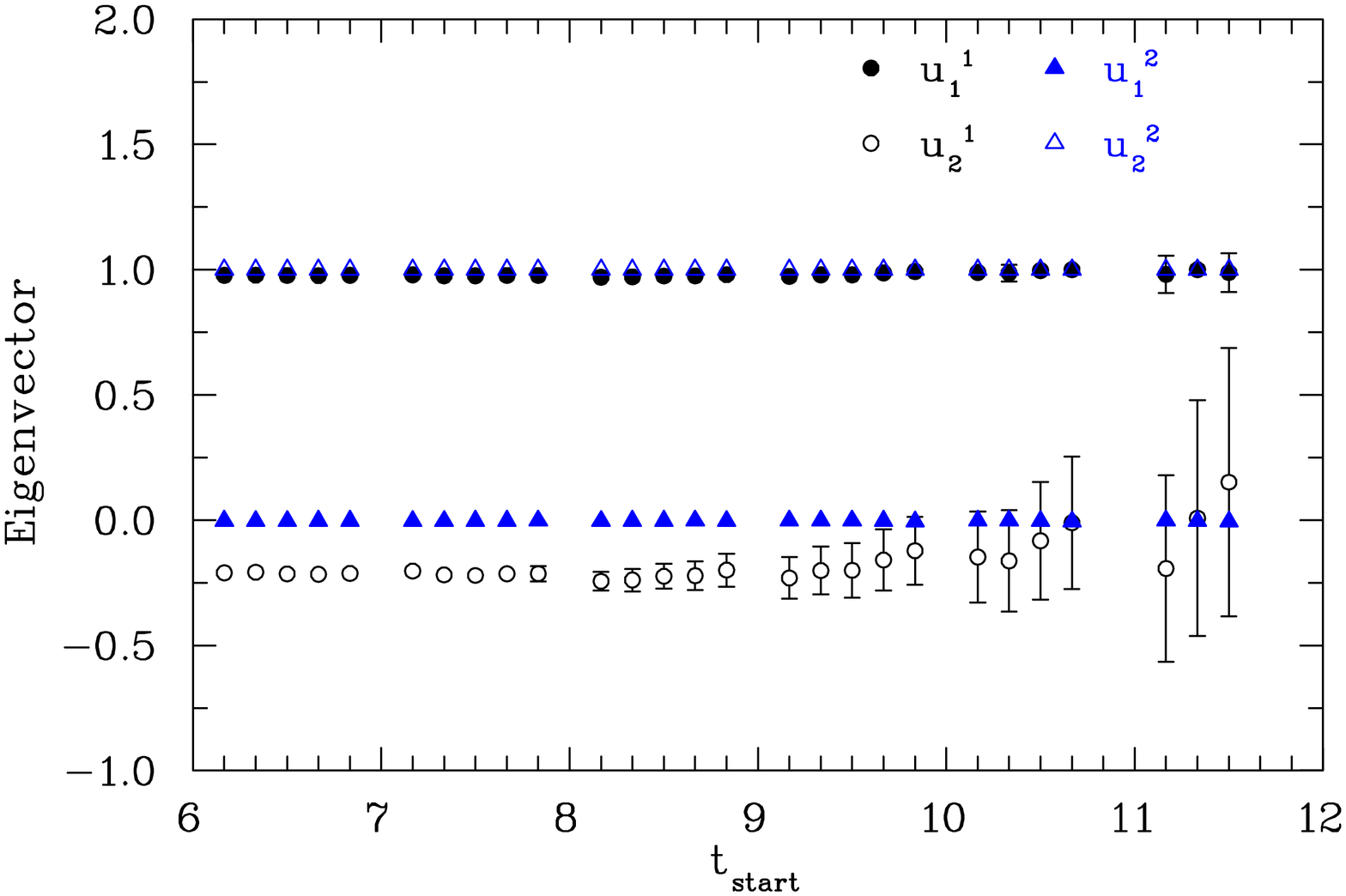}    
    \end{array}$
    \caption{(Color online). As in
      Fig.~\ref{fig:evectors_for_ptpp_for_x1x2_h1Q}, but for the
      correlation matrix analysis of
      Fig.~\ref{fig:mass_and_eig_for_so7_for_x1x2_h1Q}.}  
   \label{fig:evectors_for_so7_for_x1x2_h1Q}
  \end{center}
\end{figure*}
We now consider fits to the parity and eigenstate projected effective
mass of Eq.(\ref{eqn:efective_mass}) which follows from the
eigenvector analysis.  In this presentation, we 
consider the cut-off value for an acceptable value of
${\chi^{2}}/{\rm{dof}}$ as 1.30.  Firstly, we try to fit the effective
mass from two time slices after the source to the largest possible
time of the correlation function i.e., from time slice 6 (since the
source is at 4) to time slice 25 (after which the fixed boundary
effects are significant). We call the lower time $t_{\rm{min}}$ and
the larger time $t_{\rm{max}}$. If an acceptable fit is not obtained
then we keep $t_{\rm{min}}$ fixed and decrease $t_{\rm{max}}$ and
reattempt a fit. If this is also unsuccessful then we iterate the same
process until we reach a time near $t_{\rm{min}}$ while maintaining a
minimum fit window size.  At this point, if an acceptable fit (as
dictated by the ${\chi^{2}}/{\rm{dof}}$) is still not obtained then we
increase $t_{\rm{min}}$ by one time slice and try to fit the new
window $t_{\rm{min}}$ to $t_{\rm{max}}$. This process repeats until a
fit is obtained. The minimum fit window size we consider for the
ground state is 5 time slices in the effective mass which corresponds
to 6 time slices in $G(t)$. For the excited states, the minimum window
size of 3 is considered corresponding to 4  time slices in $G(t)$.
This provides a balance in providing evidence of an eigenstate while
avoiding residual contaminations of lower-lying states.\\
Figs.~\ref{fig:mass_and_eig_for_ptpp_for_x1x2_h1Q} and
\ref{fig:mass_and_eig_for_so7_for_x1x2_h1Q} present the ground and
excited states of the nucleon for a $2\times 2$ correlation matrix
with $\chi_{1}$ and $\chi_{2}$ interpolators for point and smeared
source correlation functions respectively. The point-like correlation
function is a difficult correlator to extract a mass from, as it
admits strong overlap with excited states. Nevertheless, we consider
the point correlation function as a challenge in this analysis.

%=================================New Figures=============================
\begin{figure*}[tph] %[!hbp]
  \begin{center}
   $\begin{array}{c@{\hspace{0.15cm}}c}  
 \includegraphics [height=0.48\textwidth,width=0.34\textwidth,angle=90]{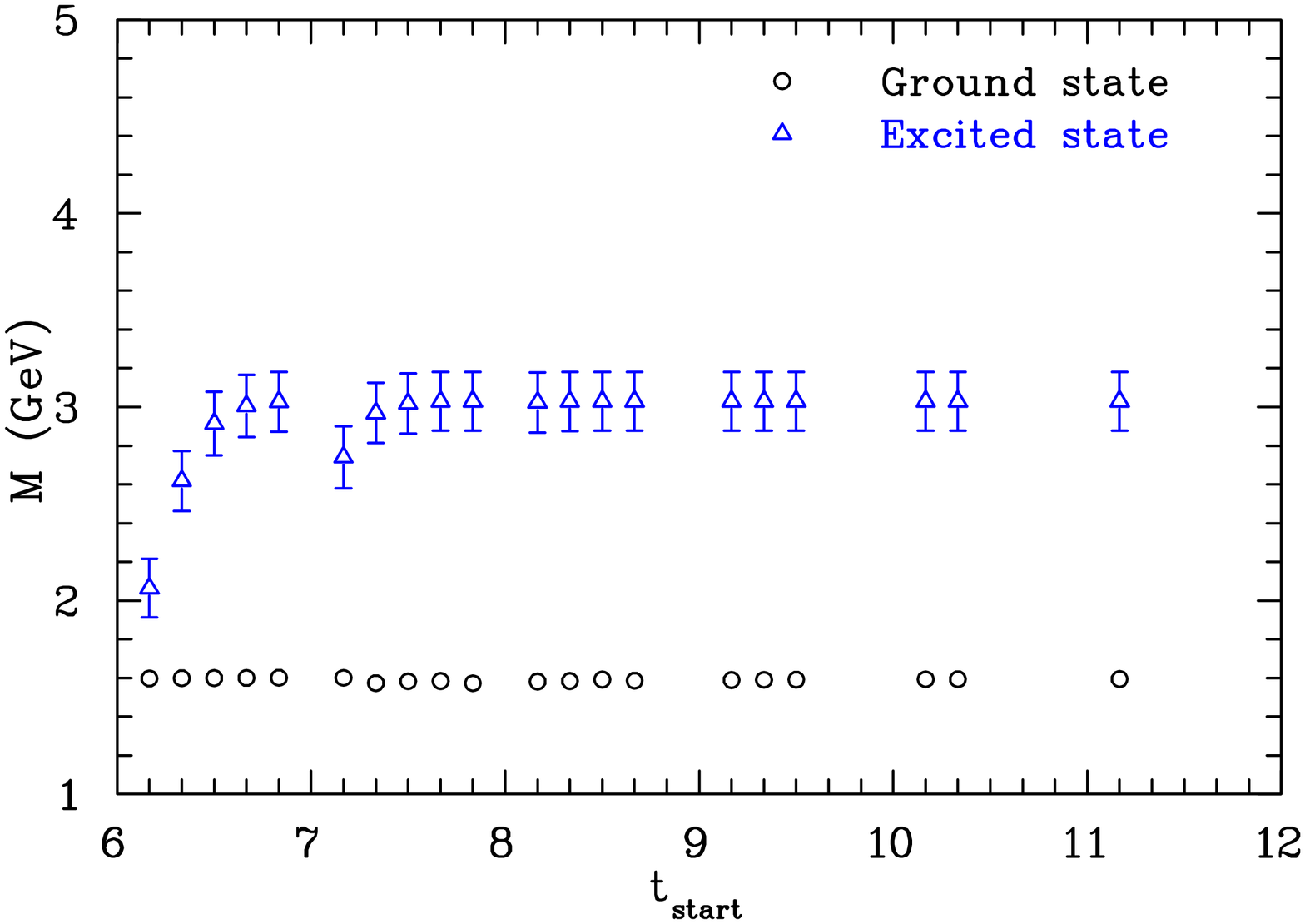} &
 \includegraphics [height=0.48\textwidth,width=0.34\textwidth,angle=90]{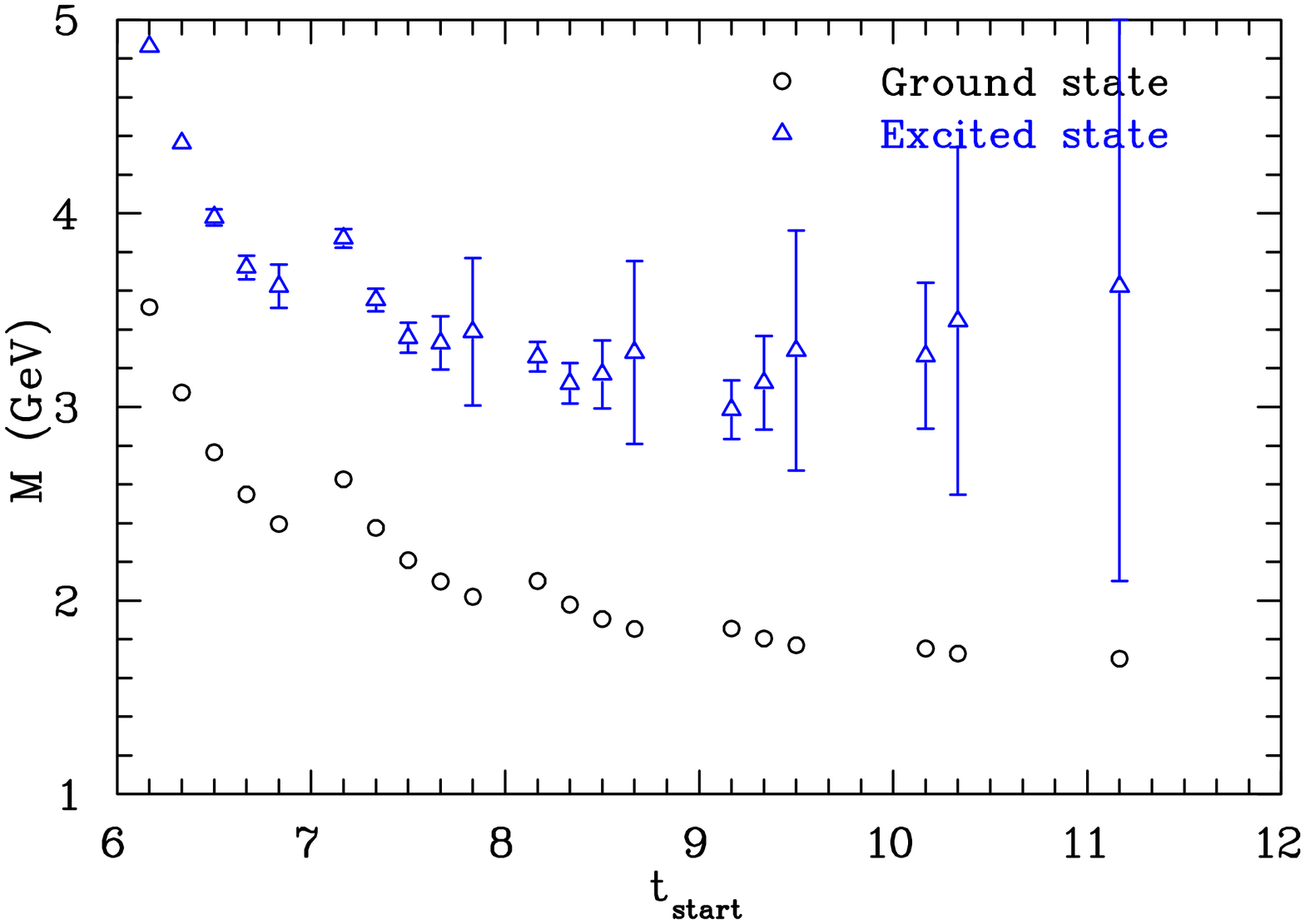}    
    \end{array}$
    \caption{(Color online). Mass of the nucleon
      ($N^{{\frac{1}{2}}^{+}}$) from the projected correlation function
      as shown in Eq.\ref{eqn:projected_cf} (left) and from the
      eigenvalue (right) for a $2\times 2$ correlation matrix of
      $\chi_{1}$ and $\chi_{4}$ interpolators. The figure corresponds
      to a pion mass of 797 MeV and for the \textbf{point source} to
      \textbf{point sink} correlation functions. Each pair of ground
      and excited states masses correspond to the diagonalization of
      the correlation matrix for each set of variational parameters
      $t_{\rm start}$ (shown in major tick marks) and $\triangle t$ (shown in minor tick marks).}  
   \label{fig:mass_and_eig_for_ptpp_for_x1x4_h1Q}
  \end{center}
\end{figure*}
\begin{figure*}[tph] %[!hbp]
  \begin{center}
   $\begin{array}{c@{\hspace{0.15cm}}c}   \includegraphics
      [height=0.48\textwidth,width=0.34\textwidth,angle=90]{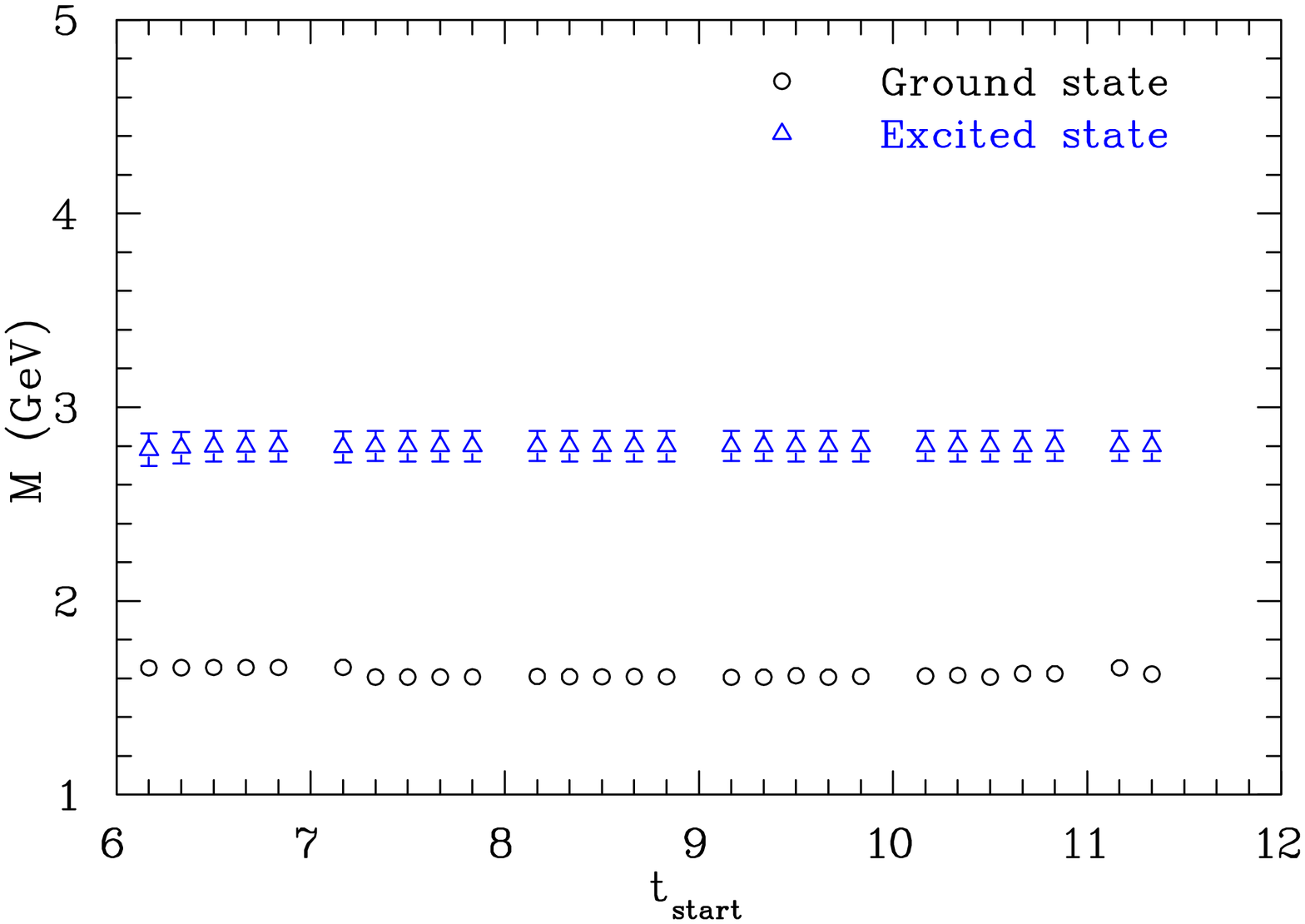}
      & \includegraphics
      [height=0.48\textwidth,width=0.34\textwidth,angle=90]{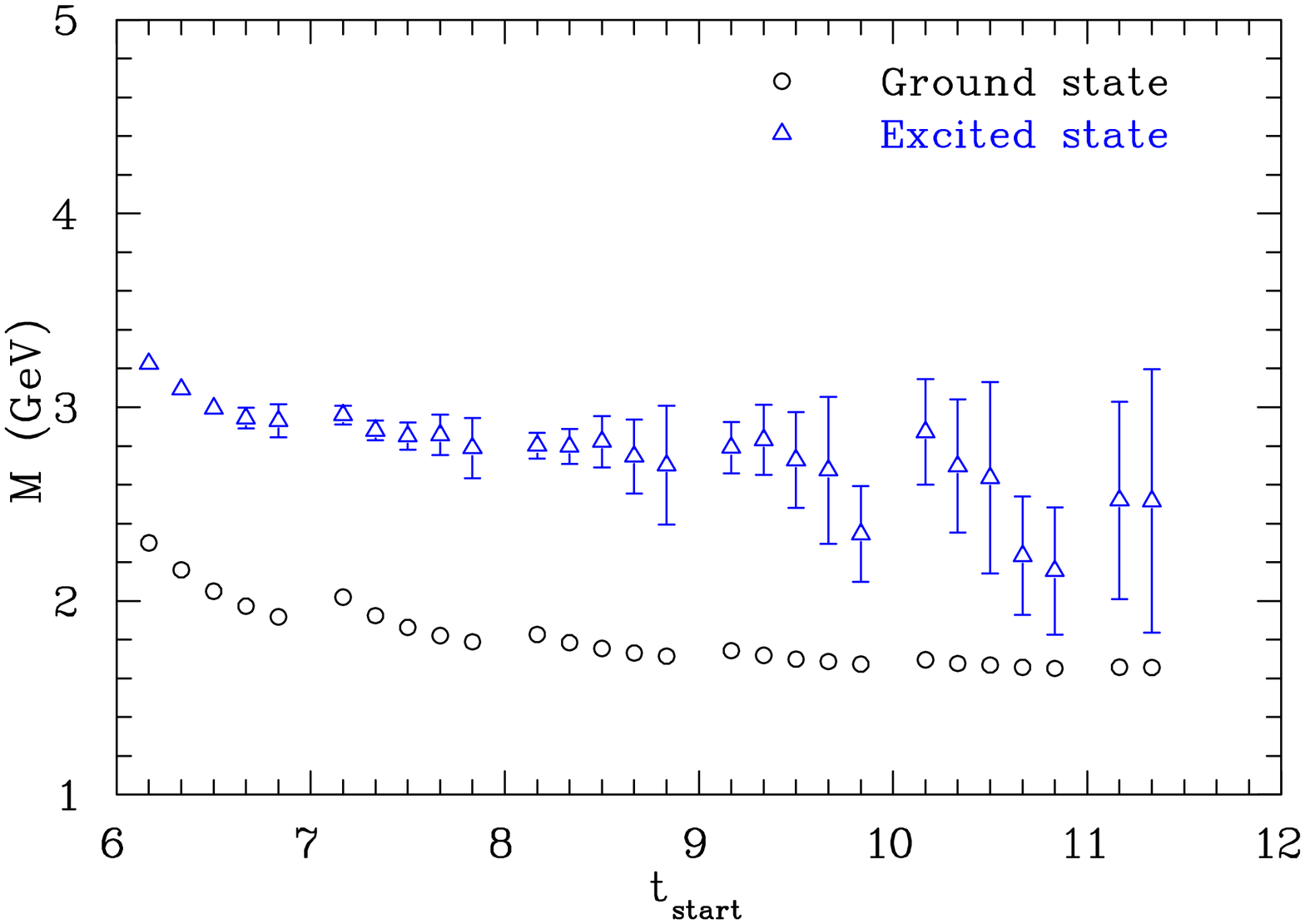}    
    \end{array}$
    \caption{(Color online). As in
      Fig.~\ref{fig:mass_and_eig_for_ptpp_for_x1x4_h1Q}, but for the
      \textbf{source smeared} to \textbf{point sink} correlation
      functions with the number of Gaussian smearing sweeps at 7, which
      corresponds to a smearing radius of 1.5831 in lattice unit.}  
   \label{fig:mass_and_eig_for_so7_for_x1x4_h1Q}
  \end{center}
\end{figure*}

The left and right figures show the mass from the projected
correlation function and the mass from the eigenvalue
respectively. Each point corresponds to the diagonalization of the
matrix for each set of variational parameters $t_{\rm start}$ and
$\triangle t$.  The mass coming from eigenvalues are intrinsic to the
variational analysis since they come directly from the diagonalization
of the matrix, while the mass from the projected correlation function
comes from the `robot' algorithm described above. It is interesting to
note that masses from the projected correlation functions are almost
independent of the variational
parameters. Fig.~\ref{fig:mass_and_eig_for_ptpp_for_x1x2_h1Q} also
shows that mass can also be extracted reliably from the point-to-point
correlation function.  The behaviour of the eigenvalues at lower
$t_{\rm start}$ and $\triangle t$ reflects the contamination of higher
excited states.  Although it can be difficult to extract a mass
directly from the eigenvalues, it is relatively easy to expose a mass
in the projected correlation function.\\
Figs.~\ref{fig:evectors_for_ptpp_for_x1x2_h1Q} and
\ref{fig:evectors_for_so7_for_x1x2_h1Q} show the eigenvectors for the
diagonalisation of correlation matrices for the point (for
Fig.~\ref{fig:mass_and_eig_for_ptpp_for_x1x2_h1Q}) and source-smeared
(for Fig.~\ref{fig:mass_and_eig_for_so7_for_x1x2_h1Q}), correlation
functions respectively.  Eigenvectors are normalized for each set of
variational parameters to unit length. It is interesting to note that
the eigenvectors do not show a strong sensitivity to excited-state
contamination.  As with the mass from the eigenvalue, at larger
$t_{\rm start}$, the eigenvectors are also dominated by
errors. Eigenvectors in Figs.~\ref{fig:evectors_for_ptpp_for_x1x2_h1Q}
and \ref{fig:evectors_for_so7_for_x1x2_h1Q} also indicate that as the
$\chi_{1}$ and $\chi_{2}$ interpolators are much orthogonal to
each other ~\cite{Melnitchouk:2002eg}, the $\chi_{1}$ interpolator has
little influence over the excited state and the $\chi_{2}$
interpolator also contributes very little to the nucleon ground state.

%=============Eigenvectors====
\begin{figure*}[tph] %[!hbp]
  \begin{center}
   $\begin{array}{c@{\hspace{0.15cm}}c}  
 \includegraphics [height=0.48\textwidth,width=0.34\textwidth,angle=90]{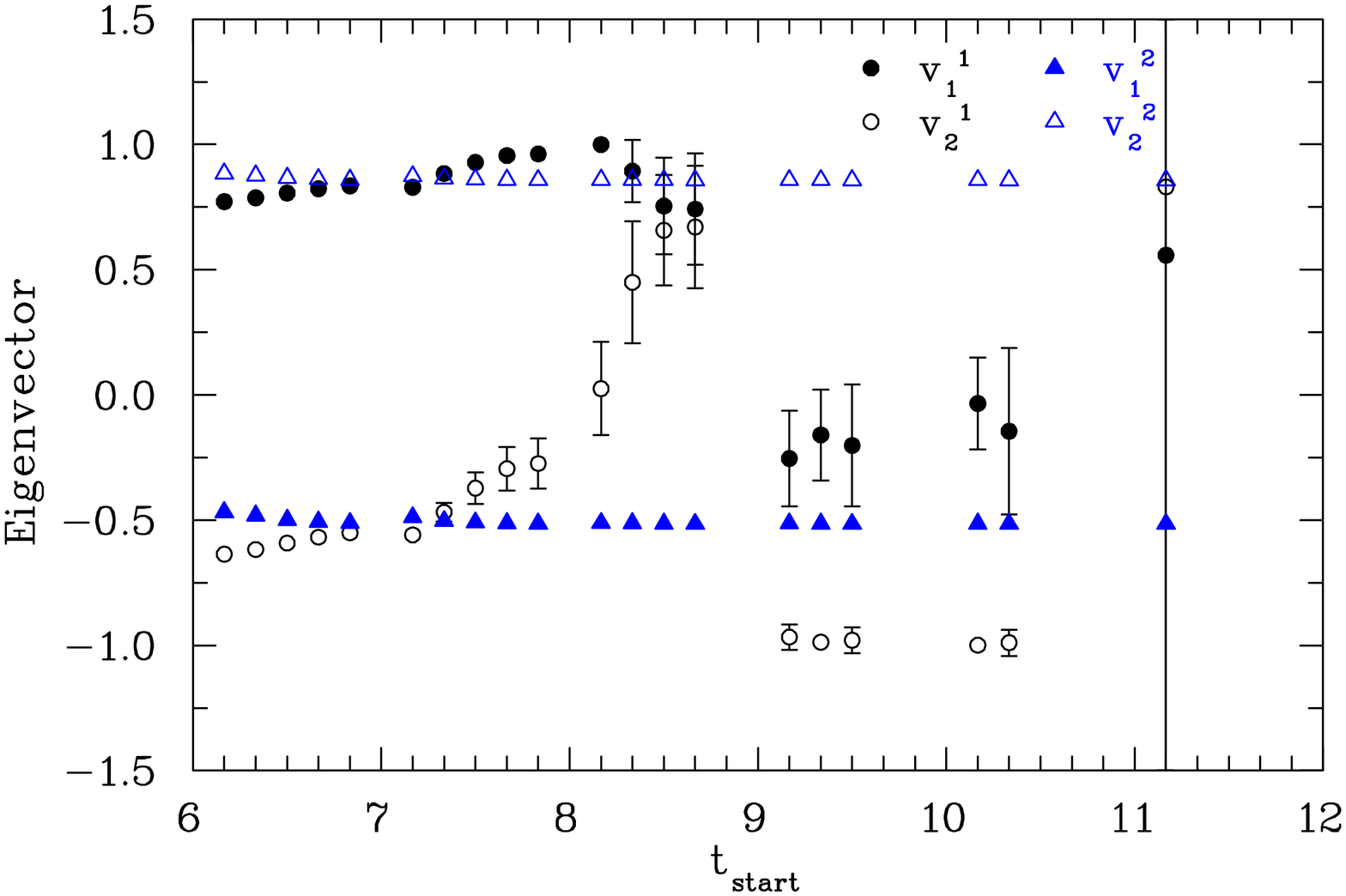} &
 \includegraphics [height=0.48\textwidth,width=0.34\textwidth,angle=90]{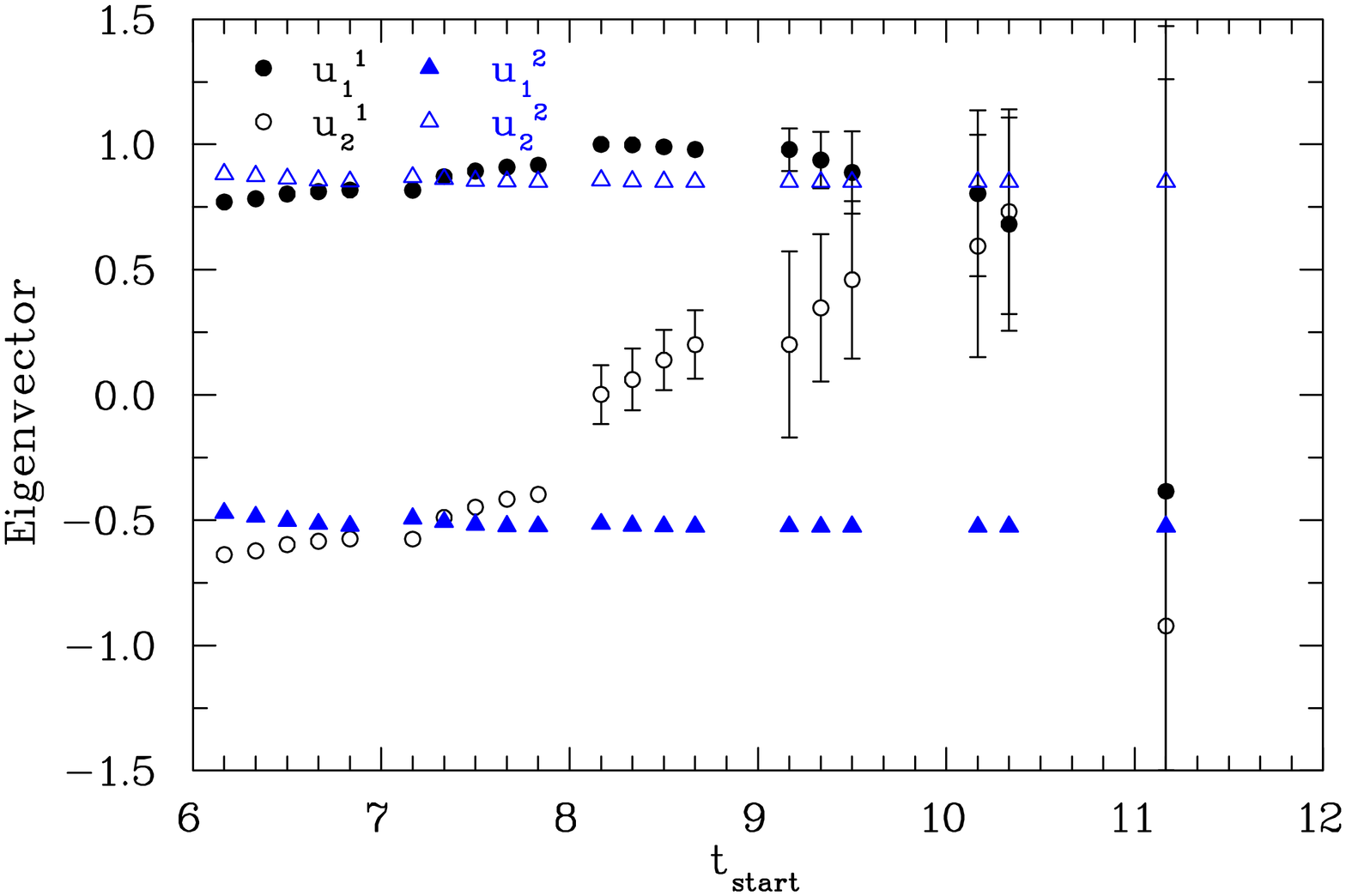}    
    \end{array}$
    \caption{(Color online). Eigenvector values of $v_{i}^{\alpha}$ (left) and
      $u_{i}^{\alpha}$ (right), as shown in
      Eq.(\ref{eqn:projected_cf}), for the 
      correlation matrix analysis of
      Fig.(\ref{fig:mass_and_eig_for_ptpp_for_x1x4_h1Q}).}  
   \label{fig:evectors_for_ptpp_for_x1x4_h1Q}
  \end{center}
\end{figure*}
\begin{figure*}[tph] %[!hbp]
  \begin{center}
   $\begin{array}{c@{\hspace{0.15cm}}c}  
 \includegraphics [height=0.48\textwidth,width=0.34\textwidth,angle=90]{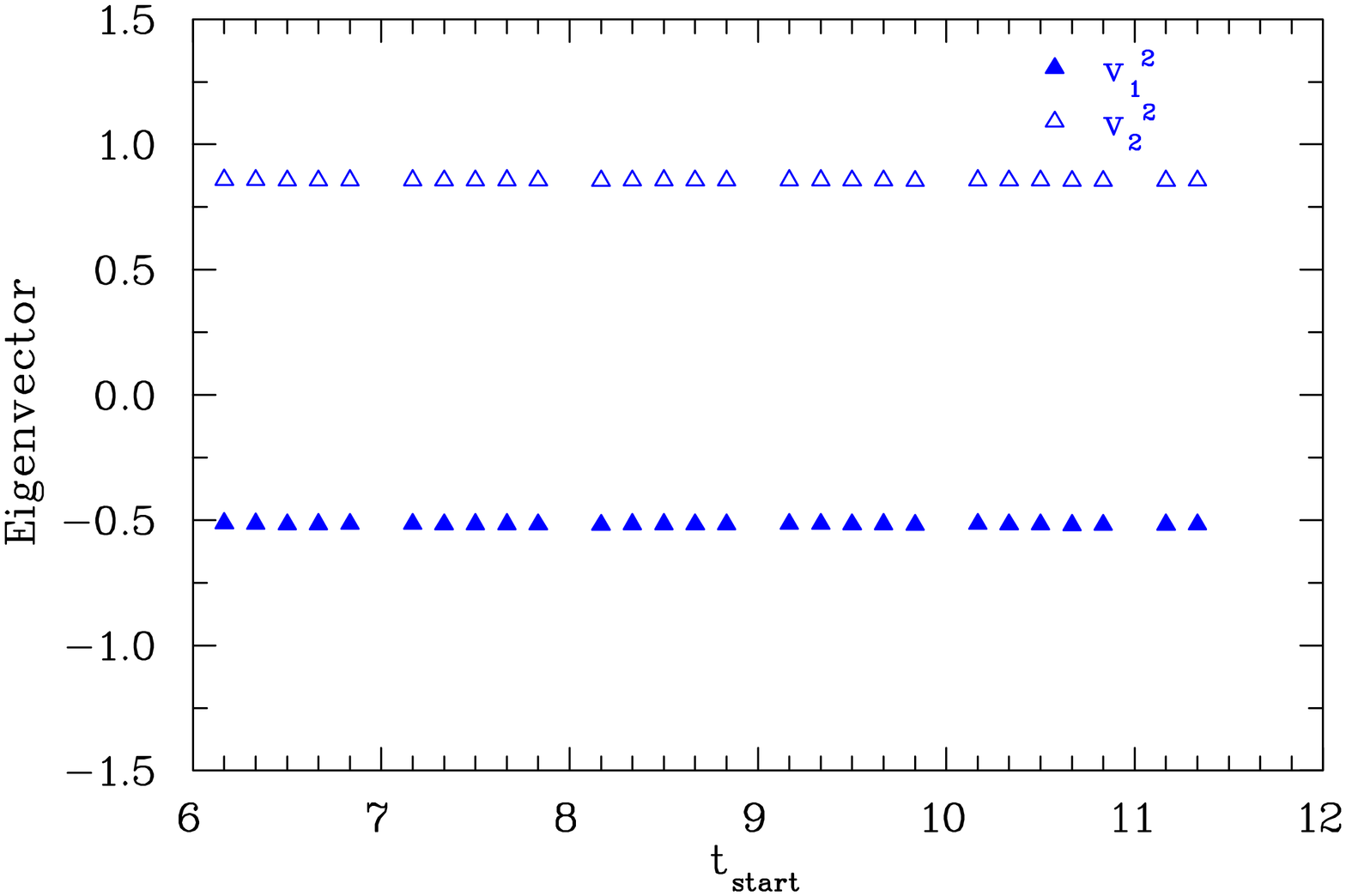} &
 \includegraphics [height=0.48\textwidth,width=0.34\textwidth,angle=90]{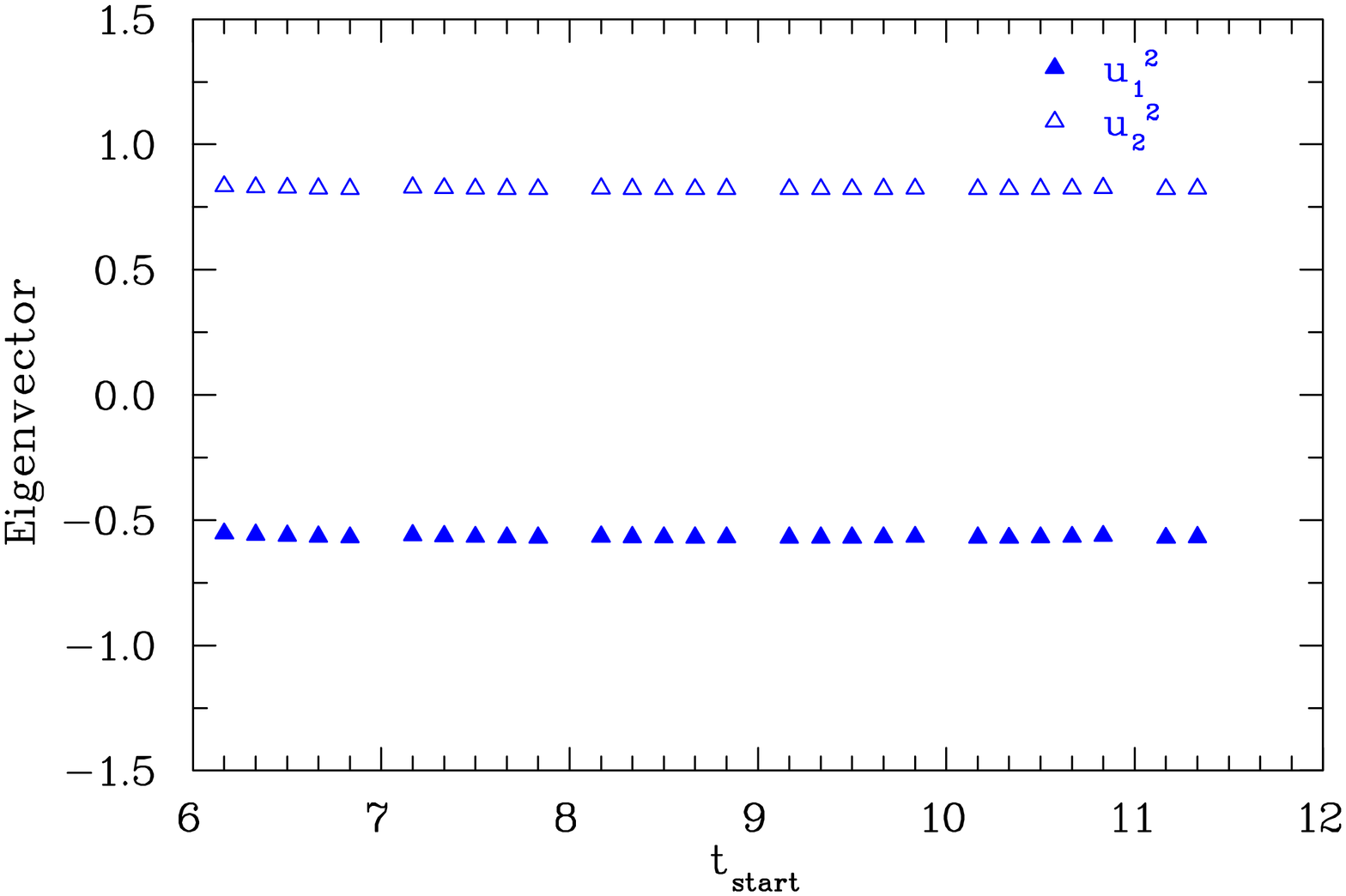}    
    \end{array}$
    \caption{(Color online). As in
      Fig.~\ref{fig:evectors_for_ptpp_for_x1x4_h1Q}, but for the
      correlation matrix analysis of
      Fig.~\ref{fig:mass_and_eig_for_so7_for_x1x4_h1Q} and for the
      excited state.}  
   \label{fig:evectors_for_so7_for_x1x4_h1Q}
  \end{center}
\end{figure*}
\begin{figure*}[tph] %[!hbp]
  \begin{center}
   $\begin{array}{c@{\hspace{0.05cm}}c@{\hspace{0.05cm}}c}  
 \includegraphics [height=0.33\textwidth,width=0.26\textwidth,angle=90]{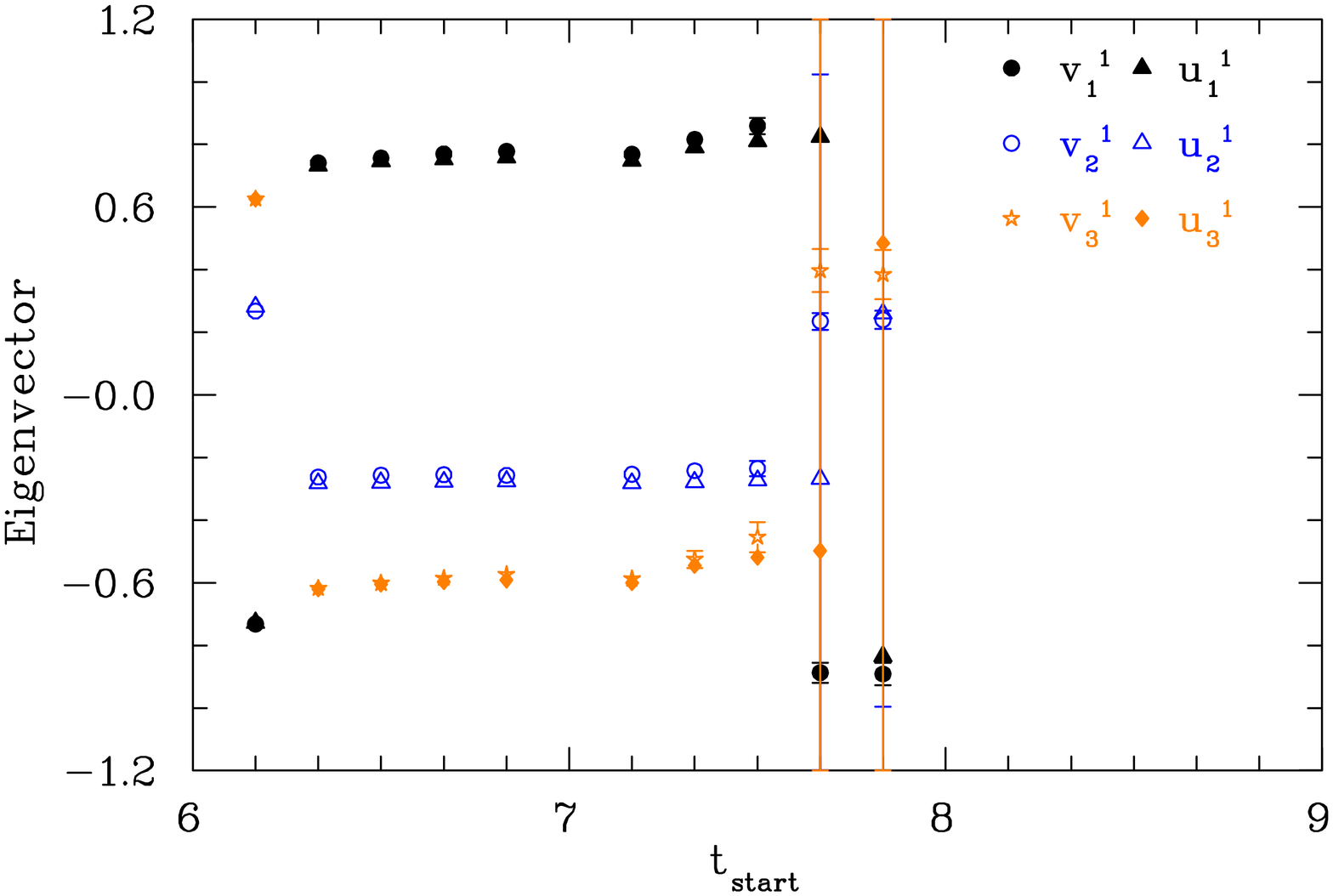} &
 \includegraphics [height=0.33\textwidth,width=0.26\textwidth,angle=90]{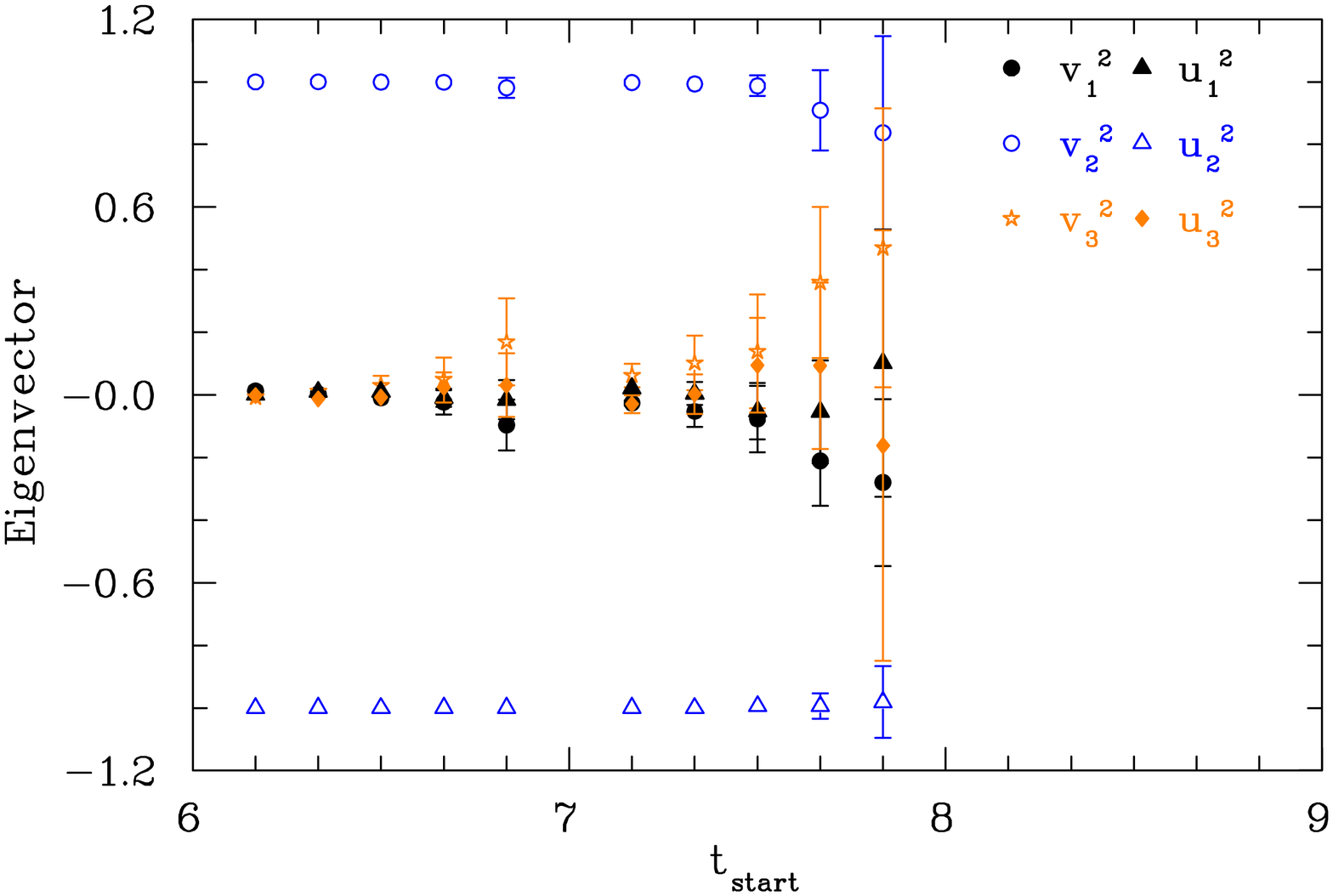} &   
 \includegraphics [height=0.33\textwidth,width=0.26\textwidth,angle=90]{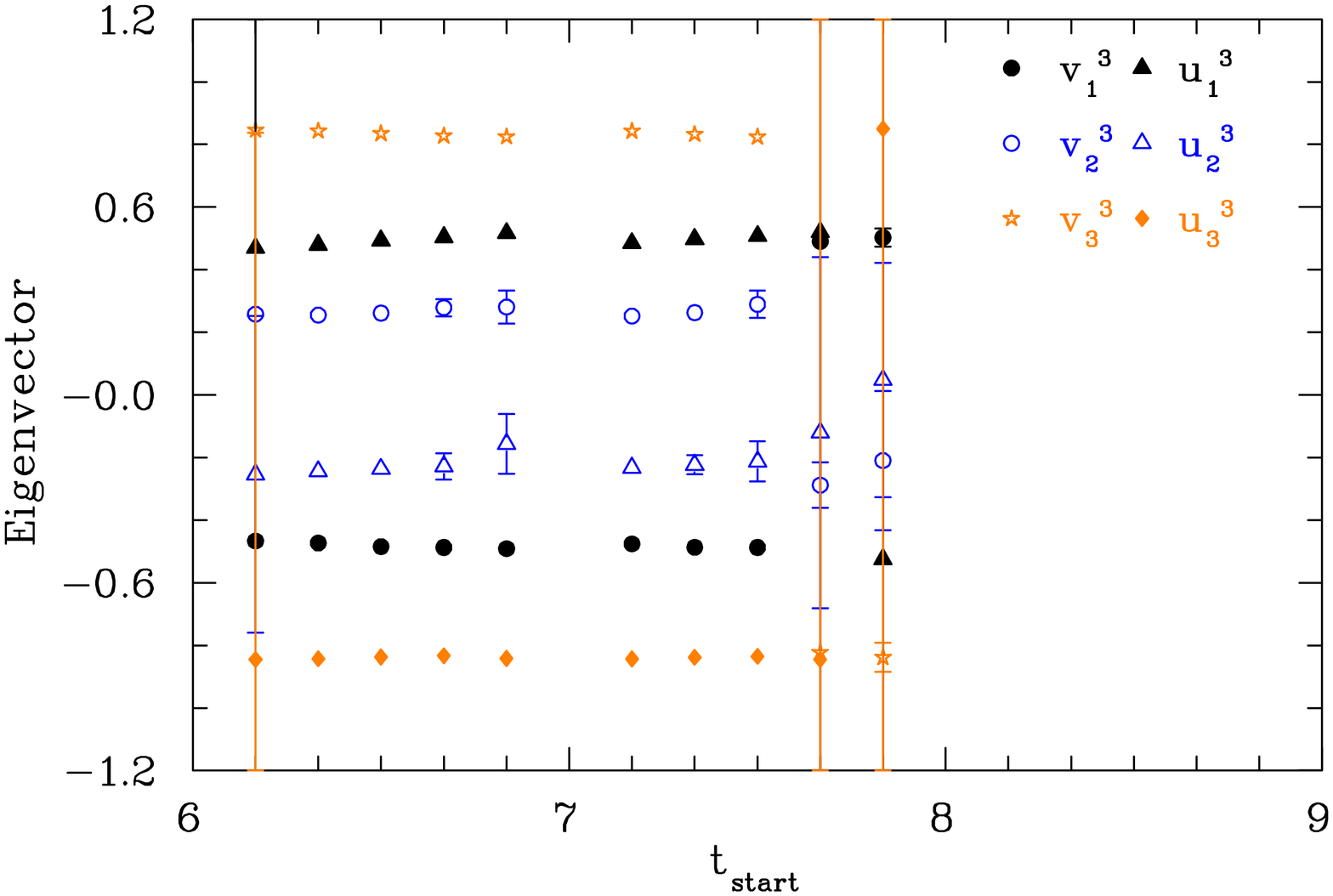}    
    \end{array}$
    \caption{(Color online). Eigenvector values of $v_{i}^{\alpha}$ and
      $u_{i}^{\alpha}$, for the $3\times 3$ correlation matrix analysis of
      $\chi_{1}$, $\chi_{2}$ and $\chi_{4}$ interpolators, for
      \textbf{point source} to \textbf{point sink} correlation
      functions and for the pion mass of 797 MeV. The left
      figure corresponds to the contributions of the interpolators to
      the ground state, while middle and right figures are for the
      first and second excited states.}  
   \label{fig:evectors_for_ptpp_for_x1x2x4_h1Q}
  \end{center}
\end{figure*}
%
%=========================================================================
%=========================================================================
%
\begin{figure*}[tph]
 \begin{center}
 $\begin{array}{c@{\hspace{0.15cm}}c}
  \includegraphics [height=0.48\textwidth,width=0.34\textwidth,angle=90]{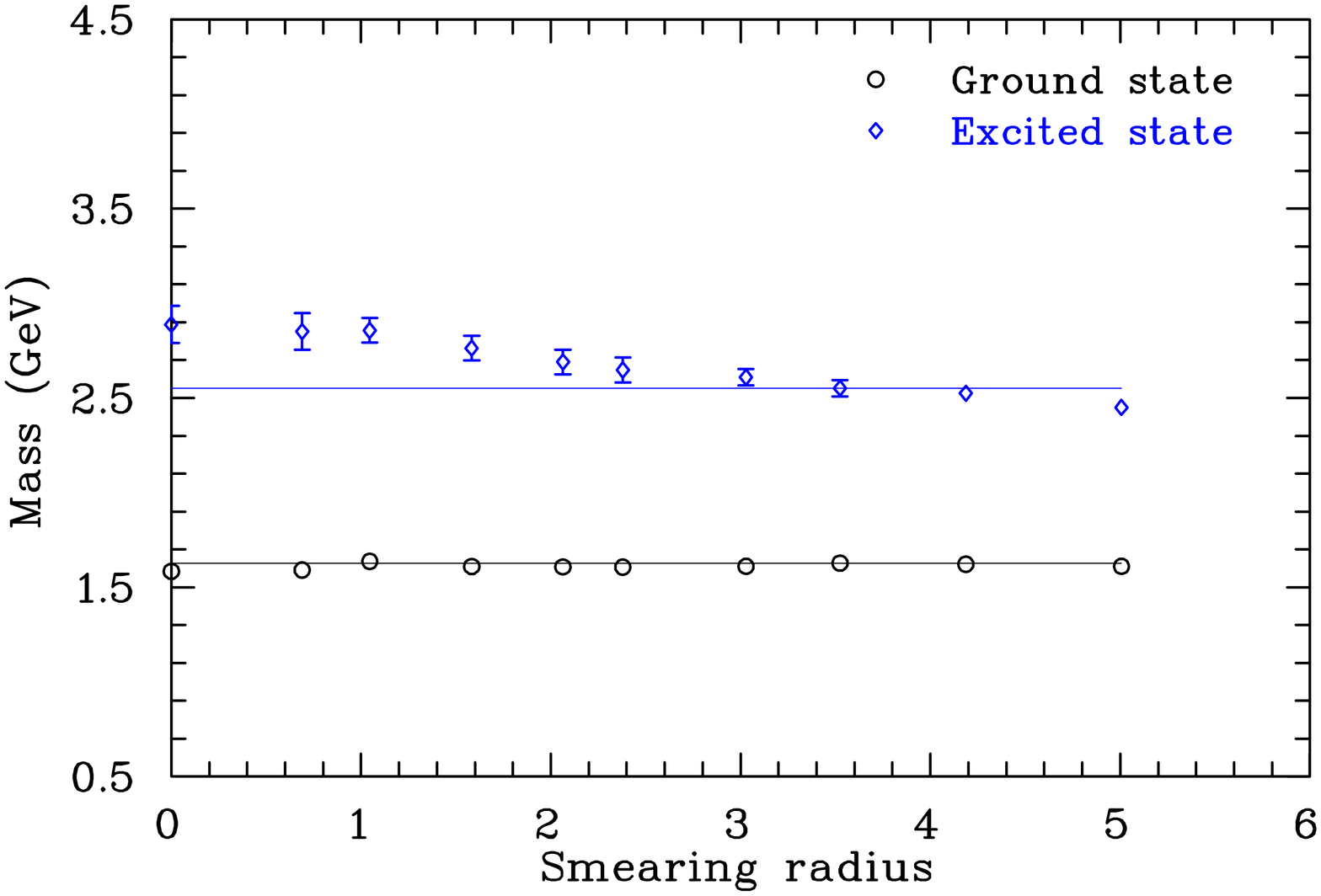} &
  \includegraphics [height=0.48\textwidth,width=0.34\textwidth,angle=90]{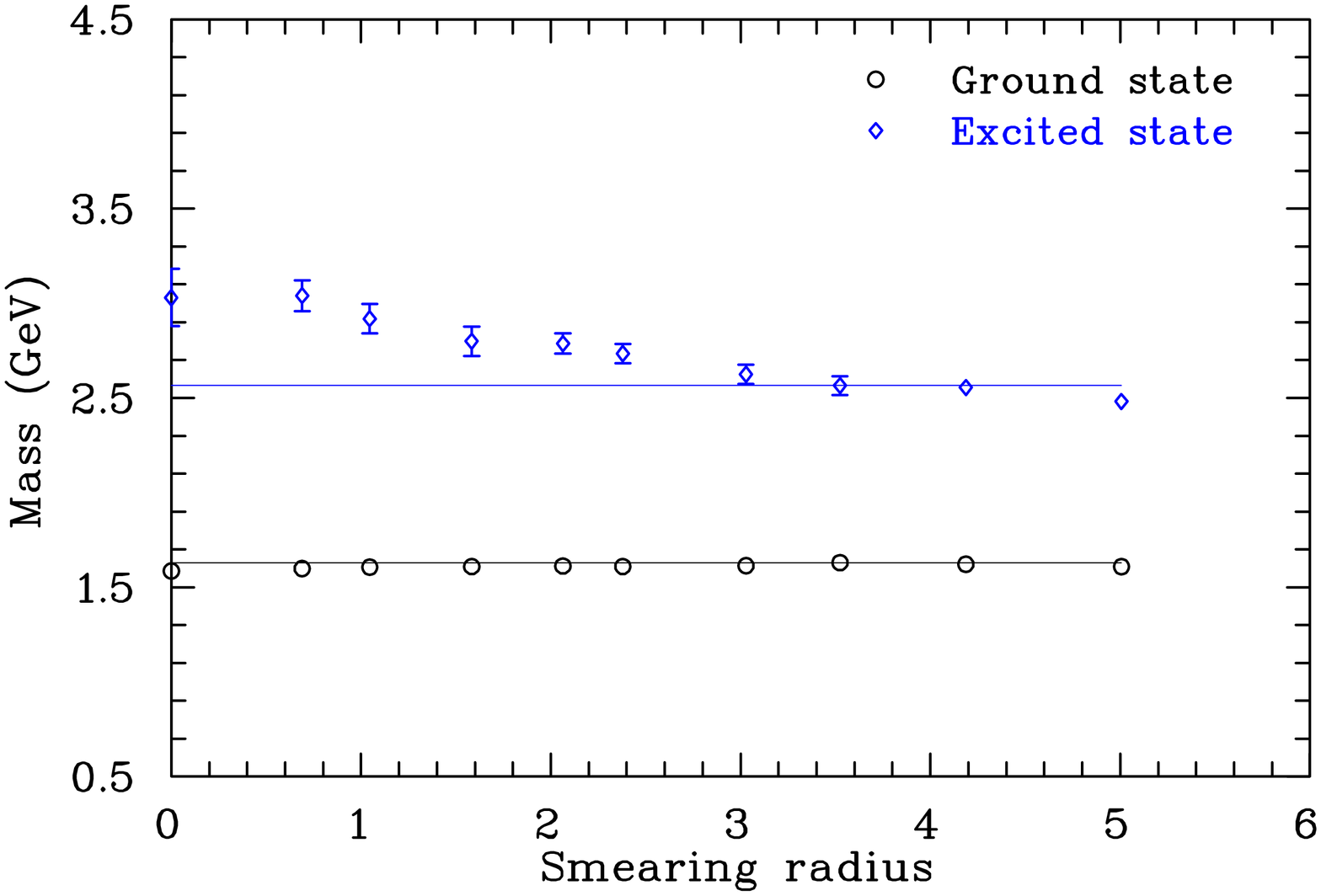}
  \end{array}$
\caption{(Color online).  Mass of the nucleon
      ($N^{{\frac{1}{2}}^{+}}$) from the projected correlation functions for
  the pion mass of 797 MeV, for \textbf{point} (leftmost point) and for
  the \textbf{source smeared} to \textbf{point sink} correlation
  functions (all other points) with rms radii 0.6897, 1.0459, 1.5831,
  2.0639, 2.3792, 3.0284, 3.5237, 4.1868, 5.0067, for $2\times 2$
  correlation matrices of $\chi_{1}$, $\chi_{2}$ (left) and
  $\chi_{1}$, $\chi_{4}$ (right).  Horizontal lines are drawn through the
  points corresponding to radius 3.5237.}  
   \label{fig:mass_for_so_all_smear_for_x1x2_x1x4_h1Q}
 \end{center}
\end{figure*}   
\begin{figure*}[tph]
 \begin{center}
 $\begin{array}{c@{\hspace{0.15cm}}c}
  \includegraphics [height=0.48\textwidth,width=0.34\textwidth,angle=90]{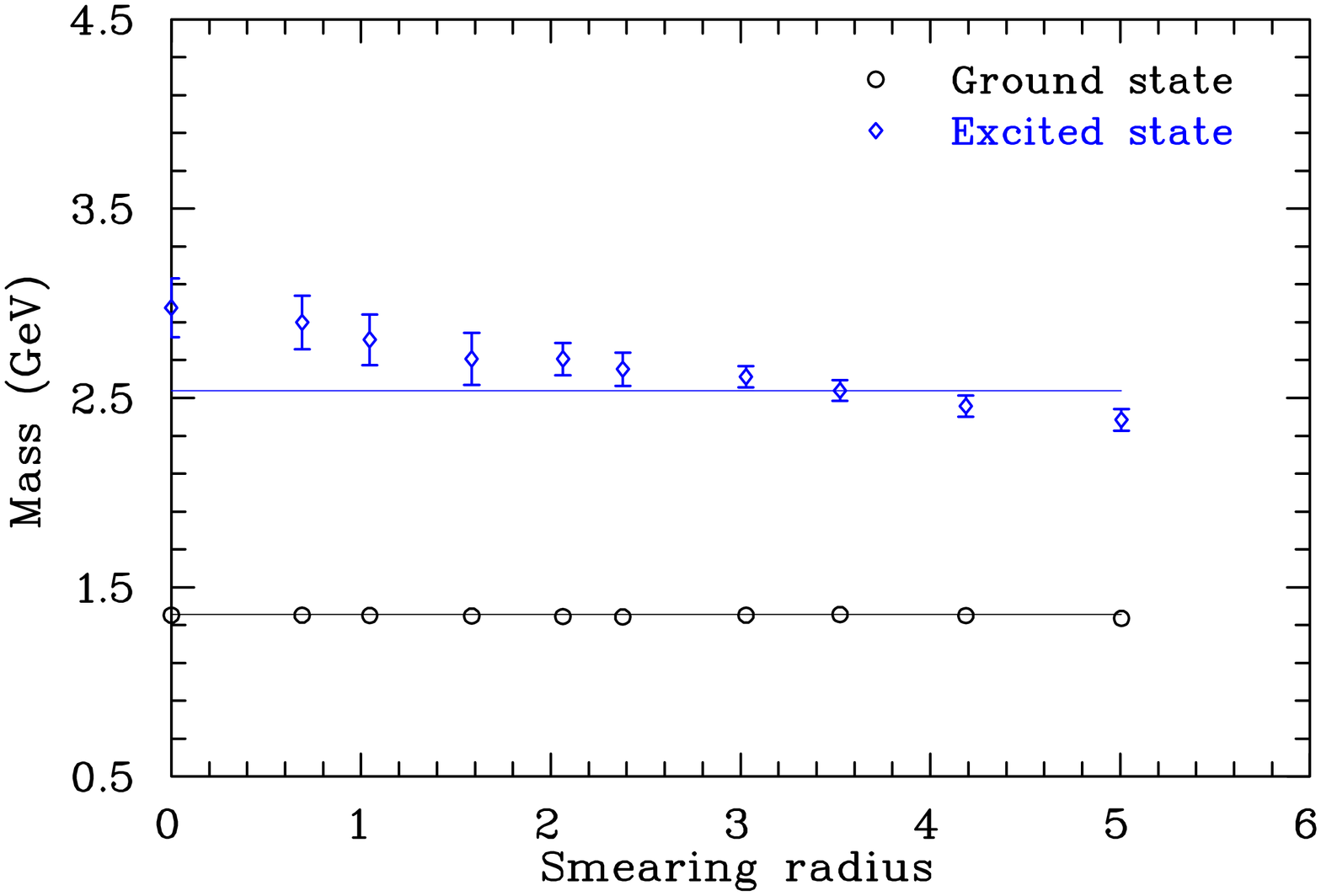} &
  \includegraphics [height=0.48\textwidth,width=0.34\textwidth,angle=90]{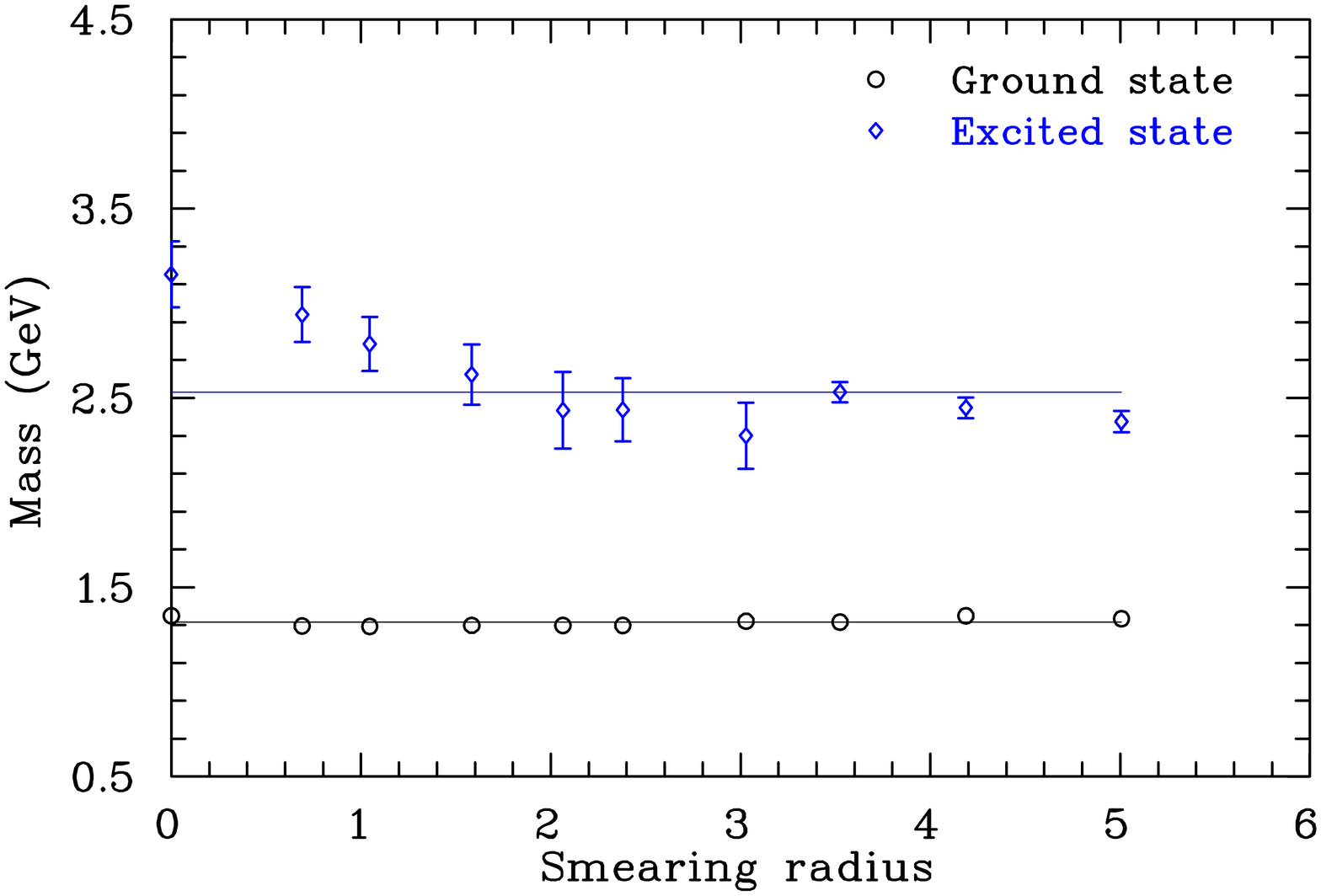}
  \end{array}$
\caption{(Color online). As in
  Fig.~\ref{fig:mass_for_so_all_smear_for_x1x2_x1x4_h1Q}, but for the
  pion mass of 541 MeV (lightest).} 
   \label{fig:mass_for_so_all_smear_for_x1x2_x1x4_h4Q}
 \end{center}
\end{figure*}   

%========================================================
\begin{figure*}[tph]
 \begin{center}
$\begin{array}{c@{\hspace{0.15cm}}c}
  \includegraphics [height=0.48\textwidth,width=0.34\textwidth,angle=90]{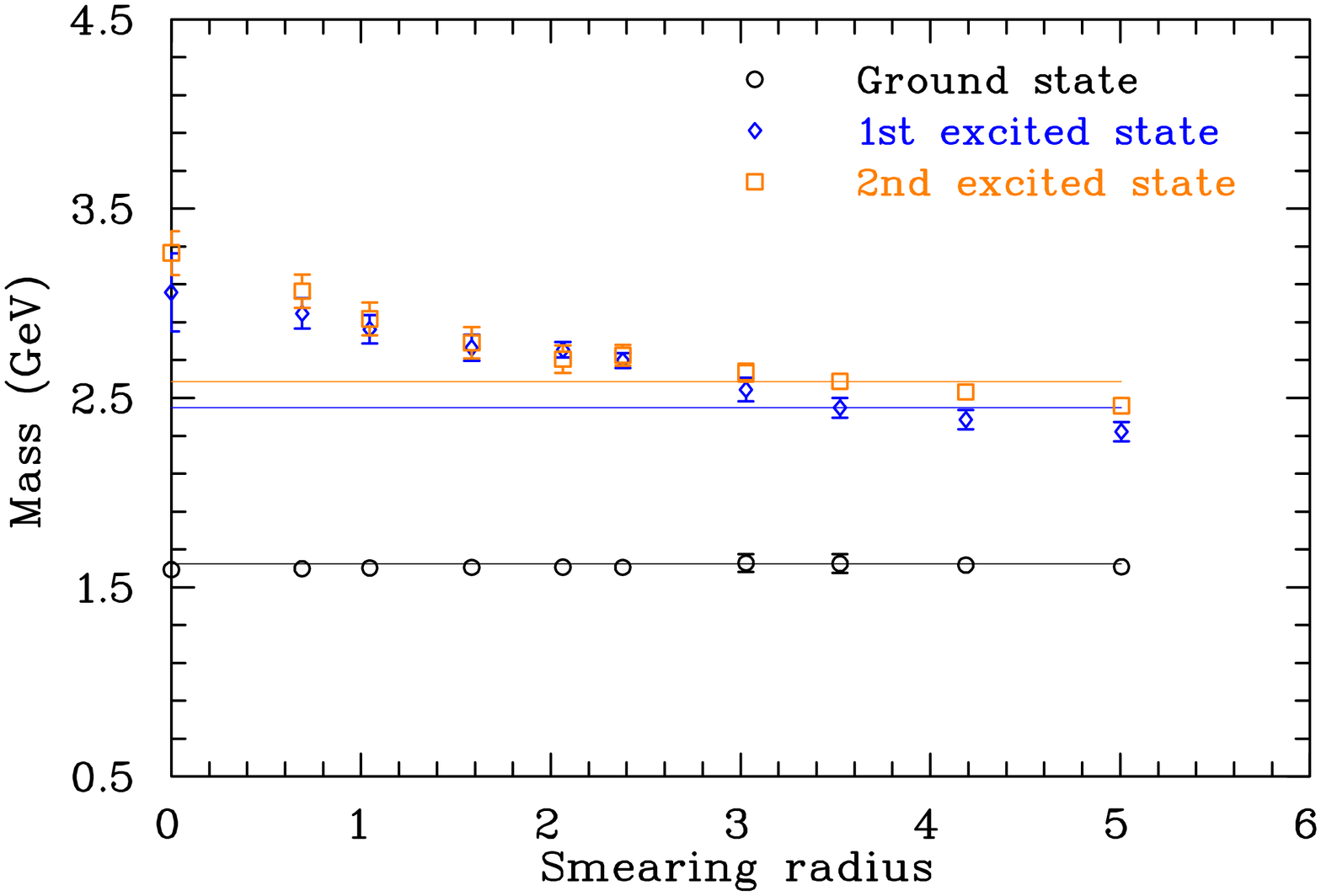} &
\includegraphics [height=0.48\textwidth,width=0.34\textwidth,angle=90]{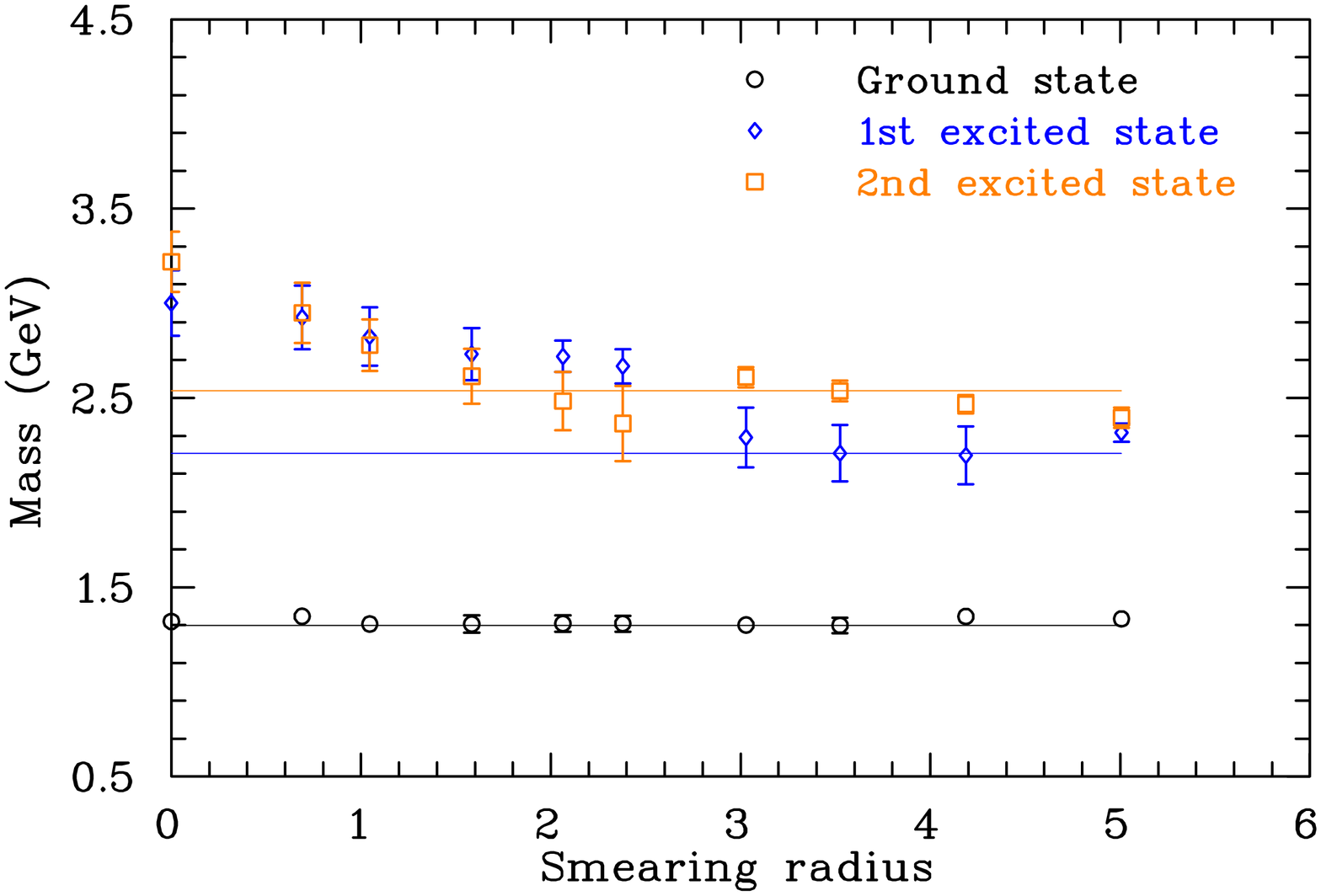}
  \end{array}$
\caption{(Color online). As in
  Fig.~\ref{fig:mass_for_so_all_smear_for_x1x2_x1x4_h1Q}, but for
  the $3\times 3$ correlation matrix of $\chi_{1}$, $\chi_{2}$ and
  $\chi_{4}$ interpolators. The left figure corresponds to the pion mass
  of 797 MeV, whereas the right figure corresponds to a 541 MeV pion mass.}  
   \label{fig:mass_for_so_all_smear_for_x1x2x4_h1-4Q}
 \end{center}
\end{figure*}

Figs.~\ref{fig:mass_and_eig_for_ptpp_for_x1x4_h1Q} and
\ref{fig:mass_and_eig_for_so7_for_x1x4_h1Q} show the ground and
excited state mass of the nucleon (as in
Figs.~\ref{fig:mass_and_eig_for_ptpp_for_x1x2_h1Q} and
\ref{fig:mass_and_eig_for_so7_for_x1x2_h1Q}) for $\chi_{1}$ and
$\chi_{4}$ interpolators. The excited state mass for the point
correlation function,
Fig.~\ref{fig:mass_and_eig_for_ptpp_for_x1x4_h1Q}, starts a little from
below and settles down after a few values of $t_{\rm start}$.  It should
be noted that similar effects also persist in the $3\times 3$
correlation matrix analysis where both $\chi_{1}$ and $\chi_{4}$ operators
are present.  However, this behaviour 
diminishes with the level of smearing as shown in
Fig.~\ref{fig:mass_and_eig_for_so7_for_x1x4_h1Q}. \\
The correlation matrix analysis of $\chi_{1}$ and $\chi_{4}$ interpolators
provide nontrivial mixing 
as illustrated in 
Figs.~\ref{fig:evectors_for_ptpp_for_x1x4_h1Q} and
\ref{fig:evectors_for_so7_for_x1x4_h1Q}.
As discussed earlier,
$\chi_{1}$ and $\chi_{4}$ are very similar and there is little to
separate them.  This is illustrated in
Fig.~\ref{fig:evectors_for_ptpp_for_x1x4_h1Q} which shows some drift
in the eigenvectors for the ground state, but with little variance in
the ground-state mass.\\
Fig.~\ref{fig:evectors_for_ptpp_for_x1x2x4_h1Q} presents
eigenvectors of the $3\times 3$ correlation matrix of
$\chi_{1}$, $\chi_{2}$ and $\chi_{4}$ interpolators for the point-to-point
correlation functions. As $\chi_{1}$ and $\chi_{4}$ interpolators
largely couple to the ground state, the left figure in
Fig.~\ref{fig:evectors_for_ptpp_for_x1x2x4_h1Q} shows the higher
contributions to the ground state come from the $\chi_{1}$ and
$\chi_{4}$ interpolators, while the first excited state (middle
figure) is completely dominated by $\chi_{2}$ interpolator and in the
second excited state (right figure) contributions from all the
interpolators are distributed. \\
To select a single mass from a series of $t_{\rm start}$ and $\triangle
t$, a value of $t_{\rm start}=8$ is preferred, and $\triangle t \geq 4$
(if possible).  We prefer to avoid the value of $\triangle t = 1$, as
it appears that $\triangle t =1$ is, in a few cases, more prone to
fluctuations than larger values.  If a mass is not obtained for these
parameters (this is the case for the lighter quark masses), then we
decrease $\triangle t$ by one time slice and try to obtain a
diagonalisation.  This procedure is repeated until the value $t_{\rm start} +
\triangle t = 10$ is reached. If the diagonalisation is still not
obtained then we decrease $t_{\rm start}$ and repeat the same procedure.
We note that our experience is in accord with that of
Ref.~\cite{Blossier:2008tx,Blossier:2009kd}. In practice we emphasise
the importance of keeping $t_{\rm start} + \triangle t$ large ~\cite{Blossier:2009kd}.

\subsection{Smearing Dependency of Excited States}
Now we discuss the smearing dependency we have observed in the masses
of the excited states. In
Figs.~\ref{fig:mass_and_eig_for_ptpp_for_x1x2_h1Q} and
\ref{fig:mass_and_eig_for_so7_for_x1x2_h1Q}, a careful comparison of
the masses for point- and source-smeared correlation functions reveals
that the excited state mass for the source-smeared case is lower in
value than for the point source.  In contrast, the ground state masses
agree within one standard deviation for almost all sets of variational
parameters.

Here we extend the analysis for various amounts of source-smearing
sweeps in the correlation functions.  Our new robust analysis
techniques reveal that the excited-state mass is smearing dependent.
In Figs.~\ref{fig:mass_for_so_all_smear_for_x1x2_x1x4_h1Q} and
\ref{fig:mass_for_so_all_smear_for_x1x2_x1x4_h4Q}, the ground-state
mass reveals no significant dependence on smearing.  However, the
masses of the excited state show a distinct systematic dependence on
the smearing radius.  Horizontal lines are drawn at the radius $\sim 3.52$
(no. of sweeps = 35) to aid in illustrating the absence of source
invariance.  Fig.~\ref{fig:mass_for_so_all_smear_for_x1x2_x1x4_h4Q}
presents similar results for the lightest quark mass considered in
this analysis.

One might search for an optimal level of smearing where the excited
state mass plateaus indicating overlap with a neighbouring state is
minimized.  However, there is no evidence of a plateau in
Figs.~\ref{fig:mass_for_so_all_smear_for_x1x2_x1x4_h1Q} and
\ref{fig:mass_for_so_all_smear_for_x1x2_x1x4_h4Q}.

This behaviour is also present for our $3\times 3$ correlation matrix
analysis as illustrated in
Fig.~\ref{fig:mass_for_so_all_smear_for_x1x2x4_h1-4Q}.  Here it is
found that the two excited states are almost degenerate and display a
similar dependence on the source smearing parameters.

Thus, we must conclude that the standard analysis of the $3\times 3$
correlation matrix of $\chi_{1}$, $\chi_{2}$ and $\chi_{4}$
interpolators is insufficient to isolate the energy eigenstates.  
The first excited-state mass revealed here is due to a linear
combination of mass eigenstates and therefore is likely to sit high
relative to the first excited eigenstate mass.

\begin{figure}[tph]
 \begin{center}
  \includegraphics [height=0.48\textwidth,width=0.34\textwidth,angle=90]{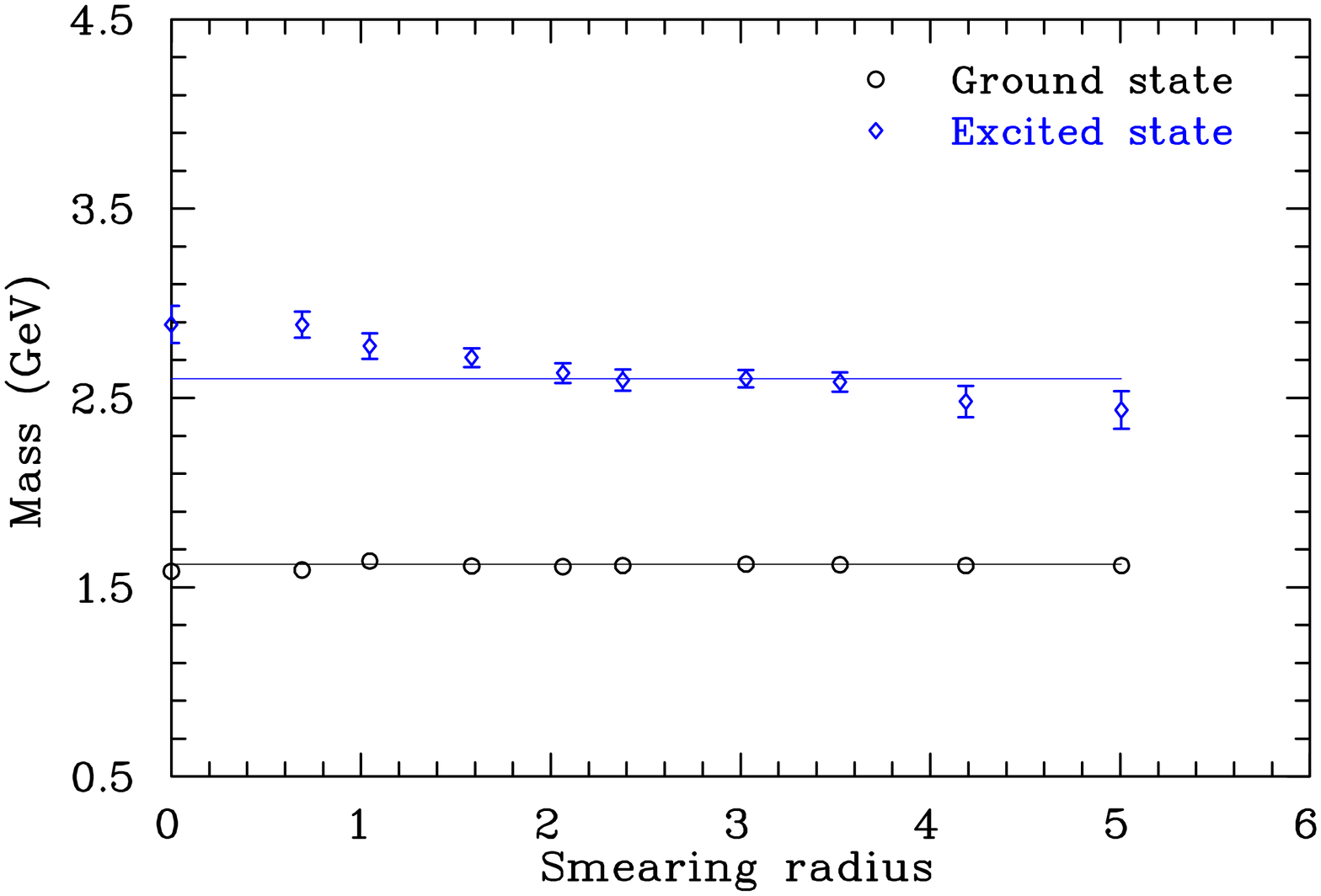}
\caption{(Color online). Mass of the nucleon
      ($N^{{\frac{1}{2}}^{+}}$) from the projected correlation functions
  for the pion mass of 797 MeV, for \textbf{point} (leftmost point) and for
  \textbf{smeared-smeared} correlation functions (all other points)
  with rms radii 0.6897, 1.0459, 1.5831, 2.0639, 2.3792, 3.0284,
  3.5237, 4.1868, 5.0067, for $2\times 2$ correlation matrices of
  $\chi_{1}$,$\chi_{2}$ interpolators. Straight lines are drawn
  through the points corresponding to radius 3.0284.}
   \label{fig:mass_for_sosi_all_smear_for_x1x2_h1Q}
 \end{center}
\end{figure}

\begin{figure}[tph]
 \begin{center}
  \includegraphics [height=0.48\textwidth,width=0.34\textwidth,angle=90]{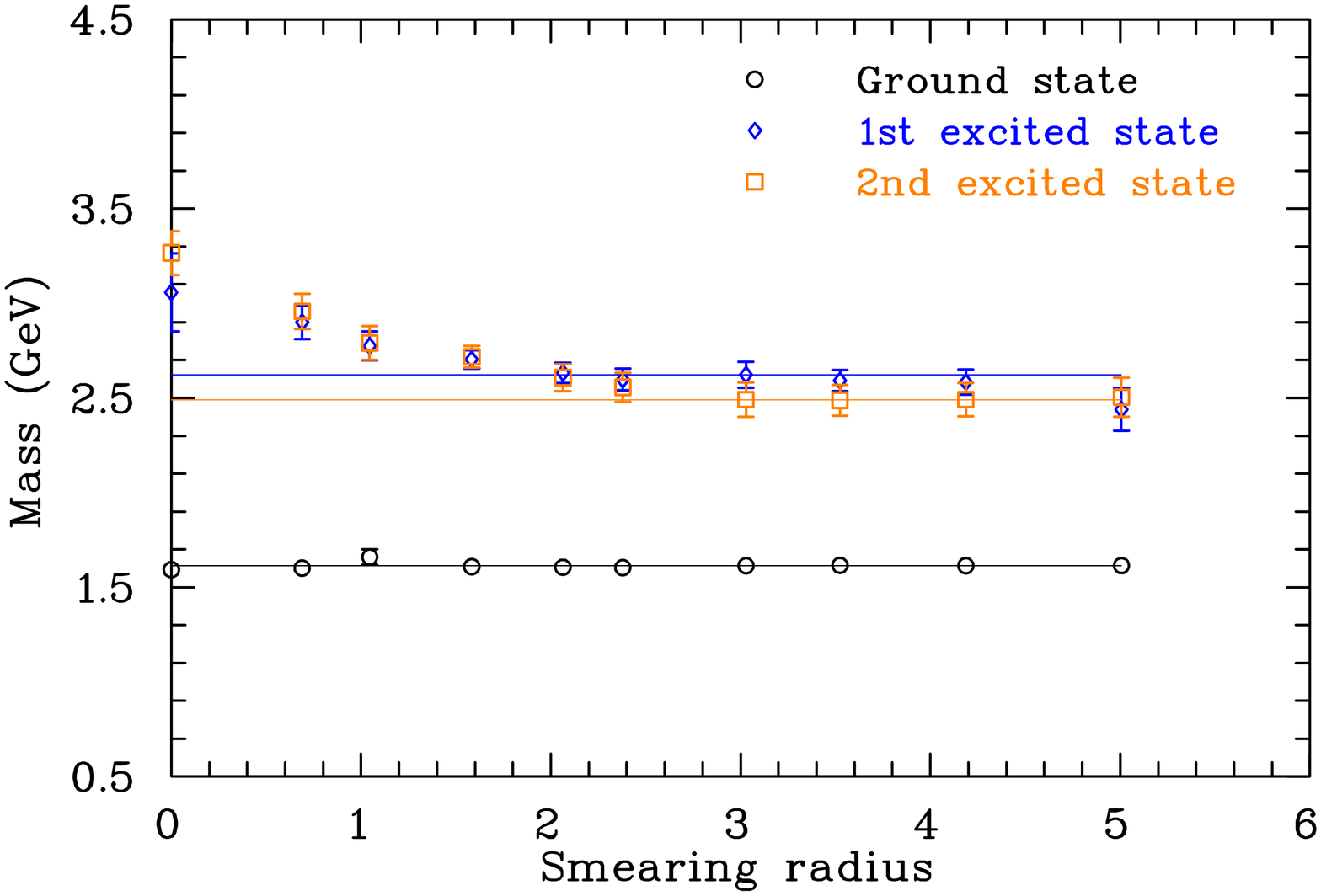}
\caption{(Color online). As in
  Fig.~\ref{fig:mass_for_sosi_all_smear_for_x1x2_h1Q}, but for the
  $3\times 3$ correlation matrix of $\chi_{1}$, $\chi_{2}$ and
  $\chi_{4}$ interpolators.}  
   \label{fig:mass_for_sosi_all_smear_for_3x3_h1Q}
 \end{center}
\end{figure}

In Fig.~\ref{fig:mass_for_sosi_all_smear_for_x1x2_h1Q}, we present
results for a variational analysis of smeared-smeared correlation
functions.  The result from the $3\times 3$ correlation matrix
analysis is shown in
Fig.~\ref{fig:mass_for_sosi_all_smear_for_3x3_h1Q}.  In this case we
also observe two nearly-degenerate excited states.  While there is
some suggestion of a plateau in this case, it seems unlikely to us
that the masses revealed here are true eigenstate masses.


It is important to consider the impact of the finite volume of the
   lattice on our observations.  It is well known that the eigenstate
   energies have a volume dependence and will change as one changes
   the volume ~\cite{Sasaki:2002sj,Sasaki:2003xc,Sasaki:2003vt}.  However, as we consider only one fixed
   volume, the eigenstate energies are fixed.  Thus the variation of
   the excited state mass revealed as our interpolating fields change
   can only be due to a superposition of eigenstates in the effective
   mass function.

\section{Conclusion}
\label{section:conclusion}
In this study we have defined and demonstrated a robust technique for
the analysis of correlation function matrices.  We have observed that
the eigenvectors describing the optimal overlap of interpolating
fields for isolating the first excited state are insensitive to the
parameters of the eigenvector analysis.  This approximate invariance
of the eigenvectors is in sharp contrast to the eigenvalue itself.
The latter changes significantly as the starting time and the change
in time is varied.  To create a robust technique for the extraction of
the excited state mass, we exploit the invariance of the eigenvectors
and construct an eigen-projected correlation function.  This
correlation function is analysed using standard analysis techniques.\\
To reduce human intervention in the fitting procedure, a
fitting algorithm has been developed which is governed by specific
fitting criteria based on the maximization of the fit window and the
minimization of the value of the ${{\chi}^2}/{\rm{dof}}$ while
commencing at the earliest time-slice possible.\\  
This study has shown for the first time that the excited state masses
are fermion-source smearing dependent for all three types of smearing
combinations; {\it i.e.}, the smeared source with point sink, the
point source with smeared sink and smeared-smeared combinations.  This
is a somewhat unexpected result given that the ground state mass is
independent of smearing.\\  
All our $3\times 3$ correlation matrix analyses provide two nearly
degenerate excited-state masses.  However, our concern is that these
two masses correspond to strongly mixed QCD eigenstates that our
analysis using the standard correlation matrix of $\chi_{1}$,
$\chi_{2}$ and $\chi_{4}$ interpolators is unable to resolve.\\
In particular, this technique has been used by several research
collaborations to determine the mass of the Roper resonance
\cite{Lasscock:2007ce,Mahbub:2009aa}.  Remarkably, mass estimates based on these
correlation matrix techniques tend to sit high relative to other
approaches.  This investigation provides a plausible explanation for
these discrepancies.\\
Finally, it is clear that changing the smearing level of the fermion
source and sink changes the relative overlap of the superposition of
the true eigenstates of QCD.  Thus it would be interesting to use the
robust analysis techniques presented here with large correlation
matrices built not only on the $\chi_{1}$, $\chi_{2}$ and $\chi_{4}$
interpolators but also on several levels of fermion-source and -sink
smearing.  This will be the subject of a future investigation.

\begin{acknowledgments}
We thank the NCI National Facility and eResearch SA for generous
grants of supercomputing time which have enabled this project.  This
research is supported by the Australian Research Council.
\end{acknowledgments}

\end{document}